\documentclass[aps,prb,floatfix,twocolumn,showpacs]{revtex4-1}
\usepackage{epsfig}
\usepackage{amsmath}
\usepackage{amssymb}
\usepackage{amsfonts}
\usepackage{ dsfont }

\usepackage[usenames,dvipsnames]{color}

\begin{document}


\hsize\textwidth\columnwidth\hsize\csname@twocolumnfalse\endcsname

\title{Fast Long-Distance Control of Spin Qubits by Photon Assisted Cotunneling }

\author{Peter Stano,$^{1,2}$ Jelena Klinovaja,$^3$ Floris R. Braakman,$^{3,4}$ Lieven M. K. Vandersypen,$^4$ and Daniel Loss$^{1,3}$}
\affiliation{$^1$RIKEN Center for Emergent Matter Science, Wako, Saitama, Japan\\
$^2$Institute of Physics, Slovak Academy of Sciences, 845 11 Bratislava, Slovakia\\
$^3$Department of Physics, Klingelbergstrasse 82, University of Basel, Switzerland\\
$^4$Kavli Institute of Nanoscience, TU Delft, 2600 GA Delft, The Netherlands}

\vskip1.5truecm
\begin{abstract}
We investigate theoretically the long-distance coupling and spin exchange in an array of quantum dot spin qubits in the presence of microwaves. We find that photon assisted cotunneling is boosted at resonances between photon and energies of virtually occupied excited states and show how to make it spin selective. We identify configurations that enable fast switching and spin echo sequences for efficient and non-local manipulation of spin qubits. We devise configurations in which the near-resonantly boosted cotunneling provides non-local coupling which, up to certain limit, does not diminish with distance between the manipulated dots before it decays weakly with inverse distance.
\end{abstract}


\maketitle

\section{Introduction}
Photon assisted tunneling (PAT) is an inelastic process where an electron overcomes a barrier by emission or absorption of photons.\cite{tien1963:PR} It shows up as additional peaks in the tunneling current with sidebands occurring at multiples of the photon frequency.\cite{tucker1985:RMP}
Although observed in superconductors already fifty years ago,\cite{dayem1962:PRL} it took three decades to establish it in other systems as well,\cite{guimaraes1993:PRL} including semiconductor quantum dots based on two dimensional electron gases,\cite{kouwenhoven1994:PRL, fujisawa1997:SM, fujisawa1997:JJAP, oosterkamp1997:PRL, blick1998:PRL,  wiel1999:CM, qin2001:PRB}
the structures of primary interest here.\cite{loss1998:PRA} With the help of a microwave  field, various forms of charge,\cite{geerligs1990:PRL,kouwenhoven1991:PRL} spin, \cite{watson2003:PRL} and photon\cite{kim1999:N} pumps were demonstrated. The adiabatic (parametric)\cite{brouwer1998:PRB, switkes1999:S, greentree2004:PRB, hohenester2006:OC, huneke2013:PRL} or photon-assisted\cite{bruder1994:PRL, stafford1996:PRL, stoof1996:PRB, pedersen1998:PRB,hazelzet2001:PRB} pumping has potential use in metrology, spintronics, and quantum information processing.\cite{oosterkamp1998:N, forster2014:PRL}

Rich opportunities offered by time-dependent control fields\cite{grifoni1998:PR,platero2004:PR} motivate us to investigate microwave-assisted manipulation of spin qubits. In particular, we consider here a linear array of electrically controlled quantum dots\cite{loss1998:PRA, hanson2007:RMP, kloeffel2013:ARCMP} and analyze electron transfer and spin exchange between distant (non-neighboring) quantum dots.\cite{amaha2012:PRB,delbecq2014:APL,takakura2014:APL,sanchez2014:PRL, busl2013:NN} Such non-local manipulations are higher order processes in the interdot tunneling amplitude (referred to as cotunneling), which proceed through virtually excited  dot states  (henceforth we refer to these as virtual states for brevity).\cite{braakman2013:NN}

We demonstrate that photon assisted cotunneling\cite{flensberg1997:PRB,braakman2013:NN} (PACT) opens up new possibilities due to the straightforward tunability of microwaves. 
Indeed, in contrast to standard electrostatic tuning of dot levels by gates,\cite{srinivasa2015:PRL} PACT allows one to tune to any virtual state,\cite{schreiber2011:NC,braakman2013:APL, braakman2014:PRB} not just the lowest one. 
Using Floquet theory, we find that cotunneling amplitudes get boosted near such resonances. 
Moreover, the virtual states possess spin structure (due to on-site exchange interaction) that results in spin-dependent cotunneling.
This dependence can be exploited, for example for spin-charge conversion, and further be tailored by spin echo protocols. Finally, tuning to a Bloch band of delocalized virtual states produced by an array of coupled dots generates exceptionally long-ranged interactions that enables coupling between distant spin qubits. Overall, we demonstrate that
PACT offers striking advantages over standard gate control of spin qubits  in terms of speed, long-range coupling, and non-locality.

\begin{figure}
\includegraphics[width=\linewidth]{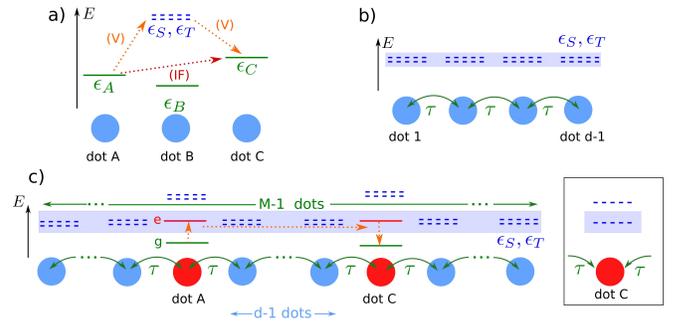}
\caption{
(Color online) 
PACT setups: a) Electrons tunnel between quantum dots A, B, and C, each singly occupied in the ``exchange'' configuration, while a single charge on the outer dots is missing in the ``tunneling'' configuration. The virtual states (dashed) are the exchange split  singlet/triplet levels $\epsilon_{S/T}$ in dot B. The photon (dotted line) is resonant with the initial-final (IF) or virtual (V) states offset.
b) An array of dots  creating a Bloch band (stripe, blue) by aligning degenerate singlet/triplet levels  (dashed, blue). c) Long distance manipulations. Dot A is singly occupied, driven, and with the excited single-electron level (e) aligned with the band (same band as in panel b). For electron transfer, dot C is empty, and gated and driven the same as dot A. For spin exchange, dot C is singly occupied, undriven, and aligned  
to the band by an exchange-split singlet level $\epsilon_{S}$ [as shown in the box]. 
}
\label{fig:scheme} 
\end{figure}

This article is organized as follows. In Sec.~II we introduce the model of the driven spin qubit array, and state and discuss the central result for the PACT amplitudes, Eq.~\eqref{eq:tac final general}. We apply it to various specific configurations in the following sections. In Sec.~III we consider PACT in a three dot structure, and point out the crucial difference between a real and virtual resonance, as only in the latter case microwaves offer a speed-up in cotunneling. In Sec.~IV we demonstrate the use of spin echos for PACT. In Sec.~V we show that, in complete analogy to the electron transfer (Sec.~III), the microwaves can boost also the non-local spin-spin exchange. In Sec.~VI we generalize to an array with many dots and demonstrate the long-distance scaling of PACT amplitudes. In Sec.~VII we analyze the errors arising during PACT, and identify their two main sources as being the incoherent leakage and the coherent charging. We show how these can be controllably limited under realistic conditions. To keep the flow of the text, we postpone detailed derivations of the main formulas to the three Appendices.

\section{Model}
We investigate PACT first on a linear array of three driven tunnel-coupled dots and subsequently extend it.
The model Hamiltonian,
\begin{equation}
H = \sum_{\alpha} \Big( H^\alpha_0 +H^\alpha_D(t) \Big) + H_T ,
\label{eq:microscopic Hamiltonian}
\end{equation}
is a sum of, respectively, single-dot confinement, driving, and inter-dot tunneling terms.
Each dot $\alpha=A,B,C$ is created by electrostatic confinement, which defines a set of single-particle dot states (with corresponding fermion creation operator $c_{\alpha i\sigma}^\dagger$) and energies, 
\begin{equation}
H^\alpha_0 = \sum_{i,\sigma} \epsilon^\alpha_{i\sigma} c_{\alpha i\sigma}^\dagger c_{\alpha i\sigma} + 
H_{\rm int}^\alpha \equiv \sum_{k} \epsilon^\alpha_{k} |k \rangle_{\alpha {\alpha}}\langle k |.
\end{equation}
Here, the orbital index $i$ and spin index $\sigma$ label the states. A uniform magnetic field $B$ sets the spin quantization axis, and enters the energies through the Zeeman term $\epsilon^\alpha_{i\sigma} = \epsilon^\alpha_{i} +\sigma g^* \mu_B B/2$, according to the electron $g$-factor $g^*$, and the Bohr magneton $\mu_B$. We neglect spin-orbit and hyperfine interactions and consider below only spin-conserving nearest-neighbor tunneling in Eq.~\eqref{eq:microscopic Hamiltonian}, which then commutes with the Zeeman term and conserves the total spin. However, we do estimate the errors originating form these neglected terms and show that they are very small for parameters of GaAs (see Sec.~\ref{sec:errors}). 

Instead of specifying the intradot Coulomb interaction $H_{\rm int}^\alpha$ for dot $\alpha$, we introduce a Fock-like basis formed by many-body states $|k\rangle_\alpha $ with different number of electrons.
Specifically, we consider states with zero ($|0\rangle$), one ($|\sigma=\uparrow,\downarrow\rangle$), and two electrons. The latter comprise a spin singlet $|S\rangle$ split by the exchange energy from the unpolarized triplet $|T_0\rangle$ and the two polarized triplets $|T_\pm\rangle$. 
Thus, $k \in \{0;\sigma; S,T_{{0},\pm}\}$. 
The array is gated such that due to charging energy costs, the doubly occupied states are relevant (and taken into account) only for the middle dot. The  total (antisymmetrized) many-body state is written as 
$|k_A\, k_B\, k_C \rangle\equiv |k l m \rangle$ with associated energy $\epsilon_{klm}$.

The interdot tunneling is described by
\begin{equation}
H_T = \sum_{\alpha\beta ij\sigma} \tau^{\alpha\beta}_{ij} c_{\alpha i\sigma}^\dagger c_{\beta j\sigma},
\label{eq:HT}
\end{equation}
where the amplitudes $\tau^{\alpha\beta}_{ij}$ are non-zero only between neighboring dots and, in general, depend on the single-particle levels they connect. 

The oscillating electrostatic potential of the dot $\alpha$ driven at frequency $\omega$ with the amplitude $V_\alpha$ shifts simultaneously  all energy levels
\begin{equation}
H^\alpha_D(t) = - \sum_{i,\sigma} e\, V_\alpha \cos(\omega t) \, c_{\alpha i\sigma}^\dagger c_{\alpha i\sigma}\, .
\label{eq:HD}
\end{equation}
Here, $e>0$ is the electron charge.
This semiclassical description of the electromagnetic field allows us to exploit the Floquet theory to derive the cotunneling amplitudes within a time-independent formalism. \cite{shirley1965:PR} We arrive at 
\begin{equation}
\begin{split}
\tau_{co}=
\sum_{\mathcal{Q},n} \frac{\langle \mathcal{P} | H_T | \mathcal{Q} \rangle \langle \mathcal{Q} | H_T | \mathcal{R} \rangle}{\epsilon_\mathcal{P}-\epsilon_\mathcal{Q} + n \hbar \omega}
 J_n\Big(\frac{eV_{\mathcal{P} \mathcal{Q}}}{\hbar\omega} \Big) J_{N-n}\Big(\frac{eV_{\mathcal{Q}\mathcal{R}}}{\hbar\omega} \Big),
\end{split}
\label{eq:tac final general}
\end{equation}
as the cotunneling amplitude between the initial state $\mathcal{P}$ and the final state $\mathcal{R}$ proceeding by virtually exciting and de-exciting a state $\mathcal{Q}$ (all being the states $|klm\rangle$). During the transition $N$ photons in total are absorbed, split to $n$ and $N-n$ at the two steps. The resonance condition 
\begin{equation}
\epsilon_\mathcal{P}+N \hbar \omega \approx \epsilon_\mathcal{R},
\label{eq:N def}
\end{equation}
defines $N$. The amplitude of an $n$-photon process is proportional to the $n$-th Bessel function $J_n$ and depends on the drop of the driving voltage amplitude between the corresponding states, $V_\mathcal{PQ}=V_\mathcal{P}-V_\mathcal{Q}$. The latter are defined as the sum of the dot driving amplitudes $V_\alpha$ weighted by the occupations $n(k_\alpha$) (the number of electrons in state $|k_\alpha\rangle$),   
\begin{equation}
V_\mathcal{P} = V_A n(k_A) + V_B n(k_B) + V_C n(k_C),
\end{equation}
assuming the many body state $\mathcal{P}=|k_A k_B k_C\rangle$.

Equation \eqref{eq:tac final general} applies to a broad range of situations, which we illustrate below on several examples. Before that, let us note that it is a pertubative result, applicable if the excited states $\mathcal{Q}$ remain virtual. Roughly, this is so if the absolute value of the first term of each summand is less than one,
limiting the photon detuning from below. Appendix A  gives a detailed discussion of this condition [see Eq.~\eqref{eqS:condition on validity}], as well as the derivation and generalization of Eq.~\eqref{eq:tac final general}. Interestingly, the same formula allows one to quantify how much the virtuality condition is broken, by introducing charging and leakage errors. The charging errors are related to the adiabaticity of the turning on/off\footnote{By changing the driving field amplitudes $V_\alpha$.} the PACT amplitudes. The leakage is characterized by a rate by which the system leaves the desired computational space, and is related to states lifetimes. To cover also the continuum model, we postpone the quantification of the errors to Sec.~VII. However, we would like to stress already here, that we observe the restrictions on the validity of Eq.~\eqref{eq:tac final general} in all cases which follow.

\section{PACT for initial-final and virtual state resonances}
 We first discuss microwave assisted {\it transfer} of electrons, which, if spin preserving, can transport spin qubits between spatially separated storage and manipulation domains.
Consider the triple dot structure containing two electrons in the "tunneling" configuration, sketched in Fig.~\ref{fig:scheme}a. The middle dot is gated below the outer dots and driven at frequency $\omega$ and amplitude $V$. We are interested in the photon assisted transfer between the outer dots, with the initial state $\mathcal{P}=|\sigma s \, 0\rangle$ and the final state $\mathcal{R}=|0\, s^\prime \sigma^\prime\rangle$, where $\sigma,\sigma^\prime,s,s^\prime=\uparrow,\downarrow$ denote spin. The virtual states $\mathcal{Q}$ then comprise the middle dot either doubly occupied or empty.

\subsection{Initial-final state resonance: the non-local PAT}

\label{sec:IFR}

We first consider an initial-final state resonance, which refers to a configuration with these states offset in energy. In analogy with the usual PAT, the $N$th sideband appears at frequency compensating the energy difference, $\epsilon_\mathcal{R}-\epsilon_\mathcal{P} \approx N \hbar \omega$. In experiments this difference is typically small compared to the charging energy. If we then neglect 
$n \hbar \omega$ in the first denominator of Eq.~\eqref{eq:tac final general}, we straightforwardly obtain [see Eq.~\eqref{eqS:tac initial final simplified}]
\begin{equation}
\tau_{co} \approx \left( \sum_{\mathcal{Q}} \frac{\langle \mathcal{P} | H_T | \mathcal{Q} \rangle \langle \mathcal{Q} | H_T | \mathcal{R} \rangle}{\epsilon-\epsilon_{\mathcal{Q}}} \right) J_N\left( \frac{eV_{\mathcal{P}\mathcal{R}}}{\hbar \omega} \right).
\label{eq:tac initial final simplified}
\end{equation}
The first bracket is the standard (photon non-assisted) cotunneling. To illustrate, for typical interdot tunneling of order tens of $\mu$eV and the charging energy of order meV (see, {\it e.g.} Tab.~I in Ref.~\onlinecite{taylor2007:PRB}), the cotunneling scale reaches order of $0.1\, \mu$eV. However, since the second term is smaller than 1 for any parameters, Eq.~\eqref{eq:tac initial final simplified} shows that the photons can only suppress the cotunneling amplitudes. It is the current across the structure which can be boosted by microwaves by aligning the states in energy at resonances.\cite{marcos2015:JAP} We conclude that in this configuration, the photons can demonstrate the existence of the non-local tunnelings, as was done in the experiment of Ref.~\onlinecite{braakman2013:NN}, but can not boost their amplitude.

\subsection{Virtual state near-resonance: cotunneling boost}

\label{sec:VR}

The situation changes dramatically if the photon is matched to virtual states. We call it a ``virtual'' resonance, defined by the initial and final states tuned to degeneracy 
electrostatically by gates, 
so that $N=0$ applies in Eq.~\eqref{eq:tac final general}, and there is a  value of $n$ and a set of states ${\mathcal{Q}}$ for which the offsets
\begin{equation}
\delta_\mathcal{ Q} =\epsilon_\mathcal{P}-\epsilon_\mathcal{ Q} + n\hbar\omega,
\label{eq:detuning definition}
\end{equation} 
are much smaller than all other offsets.\footnote{Therefore, the approximation we did to arrive at Eq.~\eqref{eq:tac initial final simplified} can not be made now.} Such terms dominate the sum in Eq.~\eqref{eq:tac final general} and other contributions can be neglected.

The condition on the states $\mathcal{Q}$  being virtual excitations, and thus on Eq.~\eqref{eq:tac final general} being valid, limits these resonances to be only near, or quasi-resonances. It means the magnitude of the offset $\delta_\mathcal{Q}$ is limited from below. Here this condition limits the maximal ratio of the photon-assisted amplitude to the non-assisted amplitude to below $eV / 2 \tau_0$. To get this estimate, we used Eq.~\eqref{eq:tac final general} for weak fields $V$, restricted the sum to a single near-resonant $\mathcal{ Q}$ state  and denoted the larger of the two matrix elements of $H_T$   as $\tau_0$. We thus obtain our first important result: for $\tau_0 \ll eV$ (weakly coupled dots), the cotunneling can be boosted by microwaves without charging the middle dot by exploiting the virtual state resonance.
 
We now analyze the  spin dependence  of cotunneling. 
We first note that the virtual state ${\mathcal{Q}}=|k 0 l\rangle$ (an empty middle dot) in Eq.~\eqref{eq:tac final general} leads to non-zero cotunneling 
only between states $\mathcal{P}$ and $\mathcal{R}$ with simultaneously $\sigma=s^\prime$ and $s=\sigma^\prime$. The lack of spin dynamics follows from the lack of any spin structure of $\mathcal{Q}$.
We thus consider the more interesting case of the photon frequency matched to doubly occupied states. To grasp qualitative features, we consider a symmetric structure with two single-electron orbitals per dot. This gives us four relevant virtual states $|0k0\rangle$ with $k$ being one of the triplets $T_{0,\pm}$ or the singlet state $S$. 
Within this simple model the three dots with two electrons are described by (see App.~B for the derivation)
\begin{equation}
\mathcal{H}_{2}=\frac{1+\eta_x}{2}\left( J \boldsymbol{\sigma}^o\cdot\boldsymbol{\sigma}^i +J_z \sigma_z^o\sigma_z^i
+ (\sigma_z^o+\sigma_z^i)b + c \right).
\label{eq:2e3d}
\end{equation}
Here the exchange energies are $J=\tau_V^2(\delta_{T_0}^{-1}-\delta_S^{-1})/2$, and
$J_z=\tau_V^2(\delta_{T_+}^{-1}+\delta_{T_-}^{-1}-2\delta_{T_0}^{-1})/2$, the effective Zeeman energy is
$b=\tau_V^2(\delta_{T_+}^{-1}-\delta_{T_-}^{-1})/2$, and the spin independent energy is
$c=\tau_V^2(\delta_S^{-1}+\delta_{T_0}^{-1}+\delta_{T_+}^{-1}+\delta_{T_-}^{-1})/2$. We denoted
$\tau_V\equiv \tau J_n(eV/\hbar \omega)$, with the nearest neighbor tunneling amplitude scale $\tau=|\tau^{AB}_{12}|$. The Pauli matrices $\boldsymbol{\eta}$ act in the pseudo-spin space with up/down being an electron in dot $A/C$. Finally, the superscripts $o$ and $i$ denote the spin of the outer ($A/C$) and inner ($B$) dot electron, respectively. 

The effective Hamiltonian $\mathcal{H}_2$ generates dynamics with spatial and spin rotations, in general, intertwined.
We first inspect symmetries by noting that $\mathcal{H}_2$ commutes with the set of operators \{$\sigma_z^o+\sigma_z^i$, $\boldsymbol{\sigma}^o\cdot\boldsymbol{\sigma}^i$, $\mathds{1}$\}$\otimes$\{ $\mathds{1}$, $\eta_x$\}. They generate, respectively, the total spin rotation around the $z$ axis and the spin-spin swap, combined either with an identity acting in the charge degrees of freedom or with a charge swap between the outer dots. The eigenvalues of these operators are therefore conserved.

\begin{figure}
\includegraphics[width=\linewidth]{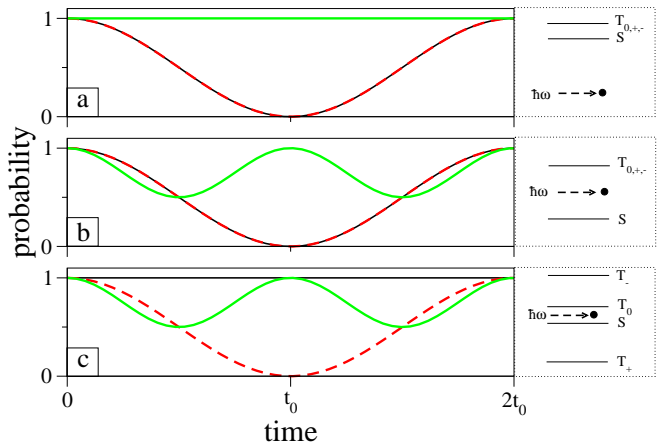}
\caption{
(Color online) 
Time evolution of probabilities for a triple dot with two electrons.  The charge on dot 
A for spin-polarized (black) and unpolarized (red dashed) electrons. The spin-down on the middle dot for unpolarized electrons (green).
The diagram on the right shows the position of the photon frequency $\hbar \omega$ relative to the offsets $\epsilon_{0 k 0}-\epsilon_{\sigma  s 0}$ of virtual states (labeled by $k$). 
}
\label{fig:dynamics} 
\end{figure}

Next, we illustrate the degree of control that microwaves offer considering special cases depicted in Fig.~\ref{fig:dynamics}. 
First, at far detuning, with all offsets approximately the same, $\delta_k\approx\delta$,
\footnote{Higher photon contributions, which might accidentally break the far-detuning condition for some $n$, are suppressed exponentially $J_n(eV/\hbar\omega)\approx (eV/2\hbar\omega)^n/n!$ at weak driving $eV\ll \hbar \omega$.}
the charge oscillates between the outer dots 
with spins frozen, see Fig.~\ref{fig:dynamics}a. This follows immediately upon noting that for such offsets the only nonzero energy in Eq.~\eqref{eq:2e3d} is $c$. Second, we take 
the microwave frequency halfway between the singlet and the triplet states, 
$\delta_{T}=\delta=-\delta_S$, which corresponds to $J=c$, and the remaining energies zero. Although spins now rotate, the charge is transferred between the outer dots with spins returned to their initial state at time $t_0=\pi\hbar\delta/2 \tau^2_V$, see Fig.~\ref{fig:dynamics}b.
Compared to the previous case, the transfer is faster since the offset $\delta$ is smaller. Third, we consider the special case $
|\delta_{T_\pm}| \gg \delta_{T_0}=\delta=-\delta_S$. This case corresponds to $J=-J_z$, while $b=0=c$, and arises if the $g$-factor is state dependent (different $g$-factor or Overhauser field for different orbital  levels) \cite{kanai2011:NN} or if the magnetic field is spatially dependent, \cite{obata2012:NJP} and the external magnetic field tunes the exchange to almost zero by orbital effects. \cite{kyriakidis2002:PRB} Alternatively, the exchange can also be tuned electrically, if the single dots of the array are replaced by singly occupied and bias-detuned double-dots\cite{petta2005:S} oriented perpendicular to the array axis.
Now, as shown on Fig.~\ref{fig:dynamics}c, at $t_0$ the charge is transferred only for unpolarized-spin configurations, effectively allowing for a single-shot measurement of the total spin by a conversion to charge. As we show below (Sec.~\ref{sec:scale}), for typical parameters the timescale $t_0$ for such photon-assisted spin manipulations can reach nanoseconds, so that PACT can outperform the purely electrostatic schemes.

\section{Spin echoes}

Spin echo techniques are standard in spin control. \cite{koppens2008:PRL,bluhm2010:N} We demonstrate their usefulness for PACT on the charge transfer discussed in the previous section (note that special configurations given in Fig.~2, identified for efficient charge transfer and spin-charge conversions, do not require any spin echoes). 
We aim at transport that is generated by the Hamiltonian $\mathcal{H}_2$ in Eq.~\eqref{eq:2e3d} but robustly spin-preserving (independently on virtual state offsets).
Such a transport corresponds to a propagator
\begin{equation}
U_{\rm eff}(t) = \exp \left ( -\frac{i}{\hbar} \mathcal{H}_{\rm eff} t \right),
\end{equation}
generated by an effective Hamiltonian 
\begin{equation}
\mathcal{H}_{\rm eff} = \frac{\tau_V^2}{4}(1+\eta_x) (\delta_S^{-1}+\delta_{T_0}^{-1}+\delta_{T_+}^{-1}+\delta_{T_-}^{-1}).
\label{eqS:echo 1}
\end{equation}
The latter is the Hamiltonian in Eq.~\eqref{eq:2e3d} with terms containing $\sigma$ matrices being removed. We find that this propagator is realized by the following echo sequence 
\begin{equation}
U_{\rm eff}(t) = U(t/4) \Pi_x^i\Pi_y^o U(t/4) \Pi_y^o U(t/4) \Pi_x^i\Pi_y^o U(t/4) \Pi_y^o,
\label{eqS:echo general}
\end{equation}
where $U(t)=\exp(-i  \mathcal{H}_2 t/\hbar)$ is the propagator generated by the Hamiltonian at hand, and $\Pi_n^{i(o)}$ is the inner(outer) spin $\pi$-rotation around axis $n$. We remind that with the echo pulses, the Zeeman term needs to be considered explicitly, as it no longer trivially factorizes from the evolution. Importantly, the echo sequences given here remove also the Zeeman term.

The general sequence of Eq.~\eqref{eqS:echo general} can be simplified for special values of the virtual state offsets. For a uniform magnetic field and a state independent $g$-factor (for which $\delta_{T_0}=\delta_{T_-}=\delta_{T_+}$), we find 
\begin{equation}
U_{\rm eff}(t) = U(t/4) \Pi_x^i U(t/4) \Pi_y^i U(t/4) \Pi_x^i U(t/4) \Pi_y^i.
\label{eqS:echo 2}
\end{equation}
Compared to the general sequence, it requires less single electron flips, which, moreover, need to be performed on a single particle only---we chose a form in which this is the inner particle (located in the middle dot) though it can be chosen as the outer dot electron as well. 

In the case of a tuned singlet-triplet degeneracy shown on Fig.~2c,  where $\delta_S=-\delta_{T_0}$ and the offsets $\delta_{T_\pm}$ are much larger, we find a sequence
\begin{equation}
U_{\rm eff}(t)=U(t/2) \Pi_z^i U(t/2) \Pi_z^i.
\end{equation}
Essentially, a single rotation suffices as the final rotation can be absorbed into the definition of the measurement at the end of the evolution.

\section{Exchange}
Next, we consider the standard setup for a spin qubit based quantum processor with single-electron dots (A, B, C).\cite{loss1998:PRA}
Can microwaves speed-up the interdot spin-spin exchange? The standard derivation for tunnel coupled dots
gives the exchange as $J_{AB}\sim 4 \tau^2 /U$ with $U$ being the single dot charging energy.\cite{loss1998:PRA} Using cotunneling amplitudes derived in the previous section, the analog of this formula gives a non-local exchange $J_{AC} \sim 4 \tau_{co}^2/U$. Such terms indeed arise and are enhanced by microwaves, but are subdominant. Namely, a resonant microwave field required to enhance $\tau_{co}$ inevitably boosts the nearest neighbor dot exchange $J_{AB}$ and $J_{BC}$.\cite{openov1999:PRB} Since these are of order $\tau^2$, the non-local exchange given above (also known as superexchange\cite{loss1998:PRA,sanchez2014:PRB})
of order $\tau_{co}^2 \sim \tau^4$ is negligible. It would then seem that the most efficient way to induce non-local operations is to concatenate nearest neighbor ones. Remarkably, we find that one can do better:\cite{lehmann2007:NN} simultaneously boosted pairwise spin rotations between dots A-B and B-C can conspire to induce a fast and useful interaction between the outer dots A and C without influencing the mediator dot B for specially tuned microwave frequency and/or if assisted by spin echoes. (For yet another alternative, see the long-distance part below).

To demonstrate this, we adopt the model described previously and, in analogy with the derivation of Eq.~\eqref{eq:2e3d}, derive the Hamiltonian 
\begin{equation}
\mathcal{H}_{3}=\mathcal{H}_{\rm SWAP} +\mathcal{H}_d+\mathcal{H}_z,
\label{eq:3e3d}
\end{equation}
describing the three dot structure gated as before, with the middle dot detuned from the aligned outer dots, but now containing three electron spins. In the above,
\begin{equation}
\mathcal{H}_{\rm SWAP} = \tau_V^2 \frac{\delta_S^{-1}-\delta_{T_0}^{-1}}{8}\sum_{\alpha=A,C} (\sigma^\alpha_+ \sigma^B_-+\sigma^\alpha_- \sigma^B_+),
\label{eq:Hswap}
\end{equation}
 implements the spin SWAP between outer dots, 
\begin{equation}
\mathcal{H}_d = -\tau_V^2 d^{-1}_+ | \uparrow \downarrow \uparrow \rangle \langle \uparrow \downarrow \uparrow |  -\tau_V^2  d^{-1}_- | \downarrow \uparrow  \downarrow \rangle \langle  \downarrow \uparrow \downarrow |,
\label{eq:Hd}
\end{equation}
describes by which the propagator departs from SWAP,
and the Zeeman-like term
\begin{equation}
\mathcal{H}_z = -\tau_V^2 \sum_s P_s E^{-1}_s,
\label{eq:Hz}
\end{equation}
is a sum of projectors $P_s$ on subspaces of total spin $z$-projection $s\in \{3/2, 1/2,-1/2,-3/2 \}$ with the corresponding inverse energies given by
\begin{equation}
E^{-1}_s\in \{2\delta^{-1}_{T_+}, 2\delta^{-1}_{T_+} + d^{-1}_+, 2\delta^{-1}_{T_-} + d^{-1}_-, 2\delta^{-1}_{T_-}\}.
\label{eq:inverse energies}
\end{equation}
We denoted $d_\pm^{-1} = (\delta_{T_0}^{-1}+\delta_S^{-1})/2-\delta_{T_\pm}^{-1}$ and $\sigma_\pm = \sigma_x \pm i \sigma_y$.
Rather than giving explicitly the cumbersome analytical results for the propagator generated by $\mathcal{H}_3$, we note that in the configuration of Fig.~\ref{fig:dynamics}c the error-causing  term $\mathcal{H}_d$ affecting the SWAP is small. Furthermore, since here $\delta_{T_+}\approx -\delta_{T_-}$ and $d_+\approx -d_-$ to a very good accuracy, flipping all spins at time $t/2$ around an axis perpendicular to the $z$-axis removes $\mathcal{H}_z$ (completely) and $\mathcal{H}_d$ (in the leading order) from the time evolution. 
We once again find that PACT qualitatively outperforms other schemes at specially designed configurations (here the one of Fig.~\ref{fig:dynamics}c).

\subsection{Scale for PACT and non-local spin-spin exchange}

\label{sec:scale}

The derivation of effective Hamiltonians $\mathcal{H}_2$ and $\mathcal{H}_3$ assumed the mediating states being occupied only virtually [condition discussed in detail around Eq.~\eqref{eqS:condition on validity}]. This requirement results in a limit on the photon offsets $\delta_\mathcal{Q}$ from below depending on the tunneling matrix element of the corresponding state.  The limit imposed on the photon detuning is $\delta \geq c \tau_V$, with $c>1$ a constant of order 1 (for example $c=3$; see Eq.~\eqref{eqS:adiabatic condition} below). Choosing typical values for lateral gated quantum dots, with tunneling $\tau_{12}^{AB}\sim 20\,\mu$eV, and the driving voltage amplitude $eV$ one order of magnitude smaller than the photon frequency $\hbar \omega$, well within the weak driving regime, we get $t_0 \geq c \pi\delta\hbar/2\tau_V^2 \sim c$ ns, the scale for photon-assisted charge and spin oscillations. Among others, this is the time scale for a spin-to-charge conversion depicted in Fig.~2c, and the analogous non-local spin-spin exchange generated by $\mathcal{H}_{\rm SWAP}$.

\subsection{Estimation of errors from neglected terms}

\label{sec:errors}

Let us now look at the errors with respect to the evolutions derived above, caused by effects we neglected so far: the interactions breaking the spin rotational symmetry of the Hamiltonian, and the presence of additional states in the Hilbert space.

First of all, we note that spin-orbit interactions do not pose a serious problem in materials where their effects are perturbative. Indeed, if the spin-orbit length $l_{so}$ is much larger than the  lateral dot scale $l_0$, in symmetric dots a spin-flip during the tunneling is extraordinarily rare, suppressed by factor $(l_0/l_{so})^3$, see Eqs.~(35), (40), and (43) in Ref.~\onlinecite{stano2005:PRB}. For GaAs dots with typical parameters $l_{so}\sim1-10\,\mu$m and $l_0\sim30-100$ nm, the spin-orbit interactions therefore lead to corrections to effective Hamiltonians in Eqs.~\eqref{eq:2e3d} and \eqref{eq:3e3d} of relative weight below 10$^{-3}$.

Second, in magnetic fields of order Tesla, which are typical for experiments with spin qubits, the effects of hyperfine spins on tunnelings are typically even smaller than those of the spin-orbit interactions (unless the nuclei are polarized intentionally). This follows from theoretical estimates on spin relaxation,\cite{raith2012:PRL} and experimental demonstrations in nuclear spin polarization.\cite{nichol2015:CM} Concerning PACT,  more important effects of nuclei will be the quasi static energy fluctuations of the virtual states. We account for this below by taking the virtual states with a non-zero spectral width.

Next, we consider the presence of additional states in the spectrum, denoted as $\mathcal{Q}^\prime$, which corresponds to additional, unwanted, channels for the cotunneling. An example is the virtual state $(\sigma 0 s)$, involving a different charge configuration, or $(0 S_2^\prime 0)$, involving an orbitally excited state (see Eq.~\eqref{eqS:B1c} for notation details). These states are offset by energies $\delta_{\mathcal{Q}^\prime}$, which are given by the charging and orbital excitation energy, respectively, both of order meV. The  relative weight of the unwanted channel [compared to the desired channel proceeding through state(s) $\mathcal{Q}$] is given by the ratio of the photon offsets $\delta_{\mathcal{Q}^\prime}/\delta_\mathcal{Q}$. 
To reduce it, it is beneficial to reduce the inter-dot tunneling itself, (say, to few $\mu$eV), which then allows one to reduce $\delta_\mathcal{Q}$ to a comparable value. This finally leads to very small relative weights of the unwanted channels, being of order $10^{-3}$.

From the above we conclude that the dominant error in the most useful scheme in Fig.~2c will be due to the finite value of the splitting of the polarized triplets. According to the previous estimates, the Zeeman energy in GaAs of 25 $\mu$eV/T then gives an error of order 10$^{-2}$ at a splitting corresponding to a field of a few Teslas. Importantly, as we discussed below Eq.~\eqref{eq:inverse energies}, these dominant errors can be further suppressed by a straightforward spin echo pulse.

\section{Long distance scaling}
To investigate the PACT amplitudes scaling with distance, we expand the array to $M-1$ singly occupied dots with uniform interdot tunneling as depicted in Fig.~\ref{fig:scheme}b. The structure is tuned such that there is a band of virtual states (indexed by $q$) delocalized over the whole array (the tuning is detailed in Fig.~1c) with wavefunctions and energies (see App.~C for details)
\begin{equation}
\Psi_q(j) = \sqrt{\frac{2}{M}} \sin\left(\frac{\pi q j}{M}\right), \, \epsilon_q = \epsilon_B + 4\tau \sin^2\left(\frac{\pi q}{2M}\right),
\end{equation} 
with $j$ the position in the array, the virtual states bandwidth given by the nearest neighbor tunneling $\tau$, and the integer indexes take values $q,j \in [1,M-1]$.

We now adopt the continuum limit, appropriate for $M\gg 1$. This, however, requires care to maintain
 the condition on the mediating excitation being virtual, so that Eq.~\eqref{eq:tac final general} remains valid. Namely, if the photon is tuned inside the band, Eq.~\eqref{eq:tac final general} shows spurious divergences (because $\delta_\mathcal{Q} \to 0$ for some $\mathcal{Q}$), and the result of its evaluation depends also on the order of the limits $M\to \infty$ and $\delta_\mathcal{Q} \to 0$. This unphysical behavior is removed by taking into account a finite  lifetime of the states.
We do so phenomenologically by adding an imaginary part to the energy with typical value $\gamma$ (up to 1 $\mu$eV for gated quantum dots), by which Eq.~\eqref{eq:tac final general} becomes
\begin{equation}
\tau_{co} = \tau_V^2 \sum_q \frac{\Psi_q^\dagger(j_A) \Psi_q(j_C)}{\delta - \epsilon_q+i\gamma},
\label{eq:tac long 1}
\end{equation}
with the band detuning $\delta = \epsilon_A- \epsilon_B + n \hbar \omega$, and the manipulated dot positions $j_A$, $j_C$, so that their spatial distance is $d=j_C-j_A$. Apart from extending the validity of Eq.~\eqref{eq:tac final general} for any detuning,
\footnote{We mean in the continuum limit: adding a finite imaginary part into the denominator of Eq.~\eqref{eq:tac long 1} the condition on the numerator being much smaller than the denominator is for $M\to \infty$ trivially fulfilled for each term of the summation, as the numerator is proportional to $1/M$.}
the physically motivated regularization by $\gamma$ has another great advantage: as shown in detail in App.~C [see Eq.~\eqref{eqS:FGR}], the imaginary part of Eq.~\eqref{eq:tac final general} for $\mathcal{P}=\mathcal{R}$ corresponds to the leakage $\Gamma$, the rate by which the system leaves the desired computational subspace. The beneficial and detrimental effects following from the presence of the mediating states $\mathcal{Q}$ can then be compared quantitatively as $\tau_{co}$ versus $\Gamma$, both of these having the functional form of Eq.~\eqref{eq:tac final general}.

To proceed with such comparison, we assume the manipulated dots are not too close to the array edges.
\footnote{The condition is $j_A,j_C\gtrsim (2/\pi) {\rm max} \{d_0,d\}$, with $d_0$ given below Eq.~\eqref{eq:tac long}. For dots closer to the array edges, the cotunneling is suppressed, unlike the leakage. For $j_A=1$ and $j_C=M-1$, representing the worst case, the cotunneling falls-off as $1/d^3$; see Eq.~\eqref{eqS:long discrete edge 3}.}
We can then replace the band wave functions by plane waves 
and reduce the numerator in Eq.~\eqref{eq:tac long 1} to a phase factor $e^{i\phi_q}$.
In the continuum limit we estimate (see App.~C)
\begin{equation}
\tau_{co} \sim 2 \tau J_n^2\left(\frac{eV}{\hbar \omega} \right)\sqrt{\frac{\tau}{\delta}} \times \textrm{min}  \{ 1, \frac{4}{\pi^2}\frac{d_0}{d} \},
\label{eq:tac long}
\end{equation}
with a crossover distance $d_0 = \surd{\tau/4\delta}$. Both the cotunneling amplitude and spatial range are boosted by tuning the photon energy to the band edge (decreasing $\delta$). 

The minimal allowed value for $\delta$ is set by the leakage.
With similar approximations as before we get (see the next Section for the calculation, and for the analysis of additional errors due to finite occupations of virtual states)
\begin{equation}
\hbar \Gamma \sim 2  \tau J_n^2\left(\frac{eV}{\hbar \omega} \right) \sqrt{\frac{\tau}{\delta}}\times \frac{\gamma}{\delta}.
\label{eq:leakage}
\end{equation}
We singled out the last term being the factor of suppression of the leakage with respect to the cotunneling. We get the natural result, that the cotunneling is ultimately limited by the states lifetime $\hbar/\gamma$.

The inverse distance decay, $1/d$ for $d>d_0$, originates from destructive interferences of the phases ${\phi_q}$, a general feature.\cite{averin1990:PRL,basharov2005:JETP,openov2005:S} Such interferences do not influence the incoherent leakage, which will therefore ultimately dominate at large distances. 
However, for intermediate distances, Eq.~\eqref{eq:tac long} gives the rate for a useful spin-preserving non-local electron transfer or  spin-spin exchange (depending on how the manipulated dots are gated, as described in Fig.~1c). Using $\gamma=\hbar/T_2^*$, with the inhomogeneous dephasing time typical for GaAs gated quantum dots $T_2^*=10$ ns, and $\tau=100\,\mu$eV, we get $d_0=6$ for detunings at which the cotunneling is one order of magnitude larger than the leakage. For these parameters, Eqs.~\eqref{eq:tac long}-\eqref{eq:leakage} predict that the cotunneling dominates the leakage for manipulations up to the 26th nearest neighbor. This result is a remarkable demonstration of how microwaves could enable coherent long-distance manipulations in spin qubit arrays.

\section{Leakage (incoherent) vs charging (coherent) errors}

In the previous section we have characterized the incoherent decay of information encoded in the system by the leakage rate $\Gamma$. Since a time necessary for an operation induced by cotunneling scales as $\hbar/\tau_{co}$, Eqs.~\eqref{eq:tac long} and \eqref{eq:leakage} give a very simple relation
\begin{equation}
P_{incoh} \equiv \Gamma \times (\hbar /\tau_{co}) \sim \gamma/\delta,
\label{eqS:Pin}
\end{equation}
for the probability of an incoherent error to occur during a typical non-local manipulation.

We now consider errors of different (coherent) origin, which we call charging errors. To understand how they arise, consider two levels energy split by $\Delta_q$ with a time dependent coupling $\tau_q(t)$. With the system being in the ground state initially, the coupling is switched on from zero to a finite value $\tau_q$. Assuming that the wave function did not change during the switch on, the upper state amplitude is 
\begin{equation}
a_2(t) = -ie^{-i\frac{\Delta_q}{2\hbar}t}\frac{\tau_q}{\hbar \omega_q} \sin \omega_q t,
\end{equation}
where $\hbar^2\omega_q^2=\Delta_q^2/4+\tau_q^2$. The system displays small amplitude and fast frequency oscillations resulting from the sudden change of the Hamiltonian. This results in the system being possibly found in the excited state after the coupling is switched off, an unwanted charging error. A crucial difference to the incoherent errors is that the probability of the coherent error is oscillatory (does not grow with time on longer timescales) and that the higher state coherent excitation is, in principle, reversible. Indeed, choosing the coupling switch-off time at $t=n\pi/\omega_q$, with $n$ integer, there will be no charging error.

However, if there are many excited states present, their populations can hardly be made to vanish all simultaneously. Moreover the dynamics is much more involved, and solving for the propagator is impractical. Looking at the previous example, one recognizes that the charging error probability is directly related to the adiabaticity of the coupling changes. For each virtual state for which these changes were not adiabatic we take its time averaged occupation probability,
\begin{equation}
P_2=\overline{|a_2(t)|^2} = \frac{2\tau_q^2}{\Delta_q^2+4\tau_q^2},
\label{eqS:P2}
\end{equation}
as the measure of its contribution to the coherent charging errors.
\footnote{An alternative simple derivation comes from defining the switching as energy conserving, rather than instantaneous. Then, assuming $\tau_q \ll \Delta_q$, the state energies after the switch-on are $-\tau_q^2/\Delta_q$, and $\Delta_q+\tau_q^2/\Delta_q$. With  the system being initially in the ground state and with zero couplings, the energy is preserved if at a finite coupling the occupation of the excited state is $P_2^\prime=\tau_q^2/(\Delta_q^2+2\tau_q^2)$. Considering that both, switching-on and switching-off $\tau_q$, generates the same amount of charging probability, we arrive at Eq.~\eqref{eqS:P2} within the perturbative order adopted in the derivations.} 
On the other hand, states for which the switching is adiabatic do not get populated and do not contribute to charging errors.

The adiabaticity of the evolution is characterized by the following ratio
\begin{equation}
\zeta = \frac{\langle \Psi_1 | \hbar \partial_t H(t) |\Psi_2 \rangle }{(\hbar \omega_{12})^2},
\label{eqS:adiabatic condition}
\end{equation}
with $\zeta \ll 1$ corresponding to an adiabatic evolution (system excitations being exponentially suppressed, as can be expected from the Landau-Zener theory). Assuming the microwave switching-on takes place on the time scale $\hbar/\gamma_{\rm rise}$, we have for the excited state $q$
\begin{equation}
\zeta_q \sim \frac{\tau_q \gamma_{rise}}{\Delta_q^2}.
\label{eqS:adiabatic condition 2}
\end{equation}
Charging errors are therefore suppressed by a slower  switching-on of couplings. On the other hand, if the cotunneling is to be used for a non-trivial operation, the Hamiltonian change must be non-adiabatic on this scale and therefore $\gamma_{\rm rise}$ is limited from below by $\tau_{co}$. We assume below that the switching-on time was chosen at this optimal value, $\gamma_{rise} \sim \tau_{co}$.

Considering now the case of a single excited state, for which $\tau_{co}\sim \tau_q^2/\Delta_q$, we get 
\begin{equation}
\zeta_q \sim \left( \frac{\tau_q}{\Delta_q} \right)^3.
\label{eqS:adiabatic condition 3}
\end{equation}
The charging errors are absent as long as the intermediate state offset $\Delta_q$ is limited from below by $c\tau_q$ with $c$ a constant larger than, but of the order of, 1. Importantly, observing this condition guarantees the validity of Eq.~\eqref{eq:tac final general} in schemes exploiting a single virtual state. We also get
\begin{equation}
\tau_{co}^{max} \sim \tau_q/c,
\label{eq:maximal cotunneling}
\end{equation}
for the maximally achievable cotunneling at which the charging errors are negligible. Taking into account Eq.~\eqref{eqS:Pin} we get another appealing conclusion, that in the case of a single excited state, the maximally achievable cotunneling is limited by the leakage if $\gamma\gg \tau_q$, and the charging if $\gamma \ll \tau_q$.

If many excited states contribute, resulting in a cotunneling amplitude much larger than an individual state contribution (as was the case in Sec.~V), Eq.~\eqref{eqS:P2} gives
\begin{equation}
P_{coh} \sim {\sum_q}^\prime \frac{2\tau_q^2}{\Delta_q^2+4\tau_q^2},
\label{eqS:P2 many}
\end{equation}
with the summation restricted to states $q$ for which  turning-on of the coupling  is non-adiabatic, 
\begin{equation}
\tau_q \tau_{co} \stackrel{!}{\leq}\Delta_q^2,
\label{eqS:adiabatic condition many}
\end{equation}
a condition we get from Eq.~\eqref{eqS:adiabatic condition 2}. Since the terms in Eq.~\eqref{eqS:P2 many} are positive, the condition $P_{coh} \ll 1$ also means that $\tau_q\ll\Delta_q$ for each term individually, as required for the validity of Eq.~\eqref{eq:tac final general}.

Using the notation from the previous section, we get
\begin{equation}
P_{coh} \sim 8\tau_V^2 \frac{1}{M} \sum_{q=1}^{q_c} \frac{1}{(\delta-\epsilon_q)^2+\gamma_c^2},
\end{equation}
where we replaced $\Delta_q^2 \to (\delta-\epsilon_q)^2+\gamma^2$ and denoted $\gamma_c^2=\gamma^2+4\tau_V^2/M$, and where the cut-off index $q_c$ is defined by
\begin{equation}
(\delta-\tau q_c^2)^2 = \frac{2}{\sqrt{M}} \tau_V \tau_{co},
\end{equation}
following from Eq.~\eqref{eqS:adiabatic condition many}. In the continuum limit we get 
\begin{equation}
P_{coh} \sim 8 \tau^2 J_n^2\left(\frac{eV}{\hbar \omega} \right) \left(\frac{\tau}{\gamma_c} \right)^{3/2} \int_{0}^{x_c} \frac{1}{([\delta/\gamma_c +x^2]^2+1},
\end{equation}
where $x_c^2=\epsilon_c/\gamma_c$. The integral can be evaluated in limiting cases of large/small values of its two dimensionless parameters. However, instead of doing so, we note that the condition in Eq.~\eqref{eqS:adiabatic condition many} is fulfilled already for the lowest state of the band if
\begin{equation}
\left(\frac{\tau}{\gamma_c} \right)^{5/2} \geq \frac{\pi}{\sqrt{M}} J_n^3 \left(\frac{eV}{\hbar \omega} \right).
\label{eq:charging suppressed}
\end{equation}
Since the leakage errors are independent of the factor $J_n$, as seen from Eq.~\eqref{eqS:Pin}, it is always possible to completely suppress the charging errors by working at a weak driving. The third power in the previous equation results in the fact that already moderately weak fields, e.g., $\hbar\omega = eV/20$, and short arrays, $M = 20$, allow one to completely suppress the charging errors at small enough offsets, $\delta/\tau= 0.003$, a value which allows for long-distance manipulations (see the example given at the end of Sec.~VI). Concluding, observing the condition in Eq.~\eqref{eq:charging suppressed}, the dominant errors in manipulations based on photon assisted cotunneling are due to the incoherent leakage and are characterized by the rate given in Eq.~\eqref{eq:leakage} and the probability given in Eq.~\eqref{eqS:Pin}.

\subsection*{PACT compared to electrostatic gating}

\label{sec:comparison}

The findings of this section also shed light on the qualitative difference between schemes based on microwave assisted and electrostatic cotunneling control. Namely, if the electrostatic gating is used and the protocol requires to focus\footnote{Focus on an excited state here means to nearly align it in energy with the ground state by using a gate potential.} on a higher lying virtual state (such as was here the case for an efficient long-distance coupling), it is necessary to cross lower lying states during the switch on/off of the cotunneling. This crossing is either fast, and then charging errors occur, or slow, and then leakage errors occur. On the other hand, when using microwaves to focus on a higher lying state, the required energy shift is provided by the microwave frequency. Since this frequency is fixed, so are all energy differences and therefore no spectral crossings happen. It is the amplitude of the higher lying state admixture in the ground state which can be changed continuously by changing the amplitude of the driving potential $V$. The freedom of tuning this admixture without including uncontrollable errors by spectral crossings is therefore available only in PACT schemes.

\section{Conclusions} We investigated the photon assisted cotunneling in an array of weakly coupled quantum dots. We found that microwaves may offer substantial advantages for various aspects of spin quit control, and demonstrated it on specific configurations. Overall, we showed how to use microwaves to increase operation speed, enhance control over the spin, and generate long-range interactions useful for spin quit manipulations.


\acknowledgements

We acknowledge support from the Swiss NF and NCCR QSIT,  IARPA, the Harvard Quantum Optics Center,
and the Dutch Foundation for Fundamental Research on Matter (FOM) and APVV-0808-12(QIMABOS).

\appendix

\section{Shirley technique for Floquet theory: derivation of Eq.~\eqref{eq:tac final general}}

Here we explain in detail how the cotunneling amplitudes are derived and calculated. The microscopic Hamiltonian of our system, Eq.~\eqref{eq:microscopic Hamiltonian}, is time dependent. In general, a calculation of a propagator for it is much more complicated than for a time independent one. However, since in our case the Hamiltonian contains only discrete frequencies, we can recast the time dependent problem into a time-independent one. The procedure is based on the Floquet theorem, \cite{grifoni1998:PR}
and was worked out in the excellent work of J.~Shirley, Ref.~\onlinecite{shirley1965:PR}. 
We now restate the results of this work that are of direct relevance for us and refer the reader therein for more. 

We restrict ourselves to the case of a single frequency $\omega=2\pi/T$ present in the Hamiltonian $H(t)$ acting in some Hilbert space spanned by a basis $\{|\kappa \rangle\}$. To map the time dependent problem into a time independent one, the following definitions are adopted. The basis is extended into tensor product states $|\kappa n \rangle \equiv |\kappa \rangle \otimes | n \rangle $, with $n$ taking integer values from minus to plus infinity. The state $|n\rangle$ is associated with the function $\exp( i n \omega t)$. A change of the value of this index corresponds to a change in the number of photons, with the correspondence explained in Ref.~\onlinecite{shirley1965:PR}. We will also use the word photons in this sense. A time dependent function $f(t)$ is associated with the matrix elements in the added part of the Hilbert space according to the following rule
\begin{equation}
\langle n | f(t) | m \rangle = \frac{1}{T} \int_0^T {\rm d} t \, 
e^{- i n \omega t}  f(t) e^{i m \omega t} \equiv f^{n-m},
\label{eqS:scalar product in t}
\end{equation}
with the last equality sign being a definition of the Fourier transform.

The propagator evolving the system from time $t_0$ to time $t$ in the original Hilbert space is given by [Eq.~(13) in Ref.~\onlinecite{shirley1965:PR}]
\begin{equation}
\begin{split}
U(t,t_0) &= \sum_{\kappa \lambda} |\kappa \rangle \langle \lambda| \sum_n e^{i n \omega t} \times \\
& \qquad \times \langle \kappa n | \exp \Big( -\frac{i}{\hbar} H_F(t-t_0) \Big) |\lambda 0 \rangle,
\end{split}
\label{eqS:Shirley propagator}
\end{equation}
with the expanded Hilbert space Hamiltonian defined as
\begin{equation}
H_F=-i\hbar \partial_t + H(t).
\label{eqS:Floquet Hamiltonian}
\end{equation}
Its matrix elements follow from Eq.~\eqref{eqS:scalar product in t}  as [Eq.~(10) in Ref.~\onlinecite{shirley1965:PR}]
\begin{equation}
\begin{split}
\langle \kappa n | H_F |\lambda m \rangle =
\langle \kappa | H |\lambda \rangle^{n-m} + n\hbar \omega \, \delta_{nm}\delta_{\kappa\lambda},
\end{split}
\label{eqS:Floquet Hamiltonian elements}
\end{equation}
with $\delta$ the Kronecker delta symbols. The advantage of the described mapping can be appreciated from Eq.~\eqref{eqS:Shirley propagator}, where the second line takes the form of a propagator of a time-independent problem.  Therefore, its calculation is amenable to corresponding perturbative techniques. In another words, the propagator calculation is reduced to a matrix $H_F$ eigenvalue problem. 

Let us now consider an illustrative case. Suppose the Hilbert space $\{| \kappa \rangle \}$ consists of three states  denoted as $\kappa=\mathcal{P},\mathcal{Q},\mathcal{R}$. This covers essentially all configurations considered in the main text with the three states being, respectively, an initial, virtual, and final state, upon various different identifications of these with the states $| k l m \rangle$. As a specific example, one might consider two electrons in three dots, with the initial state being the right most dot empty, $\mathcal{P}=|\sigma s 0\rangle$, the final state being the left most dot empty, $\mathcal{R}=|0 \sigma s\rangle$, and the virtual state being the middle dot empty, $\mathcal{Q}=|\sigma 0 s \rangle$, with spins $\sigma,s$ fixed.
The middle dot, driven by microwaves, is gated such that its other states (such as doubly occupied) are far away in energy making their contribution negligible. (The condition will be specified more 
precisely later). The Hamiltonian restricted to this three state subspace is
\begin{equation}
\begin{split}
H(t) & =  \sum_{\kappa=\mathcal{P},\mathcal{Q},\mathcal{R}} \big( \epsilon_\kappa +eV_\kappa \cos(\omega t) \big) |\kappa\rangle \langle \kappa |  + \\
&+ \left( \tau_{\mathcal{P}\mathcal{Q}} |\mathcal{P}\rangle \langle \mathcal{Q} | + \tau_{\mathcal{R}\mathcal{Q}}  |\mathcal{R}\rangle \langle \mathcal{Q} | 
+{\rm h.c.} 
\right), 
\end{split}
\label{eqS:H1}
\end{equation}
comprising the energies and inter-dot tunneling terms. Here, h.c. stands for Hermitian conjugate. The driving amplitude of the total many-body state $\kappa=|k_A k_B k_C\rangle$ is defined as $V_\kappa = V_A n(k_A) + V_B n(k_B) + V_C n(k_C)$ with $n(k)$ the number of electrons in the state $|k\rangle$. Our choice of driving the middle dot at potential amplitude $V$ gives $V_\mathcal{P}=V$, $V_\mathcal{Q}=0$, and $V_\mathcal{R}=V$. Finally, 
\begin{equation}
\tau_{\kappa \lambda} = \langle \kappa | H_T| \lambda\rangle,
\label{eqS:tau definition} 
\end{equation}
denotes the tunneling amplitudes.   

Next, we calculate the matrix $H_F$ with $H(t)$ given by Eq.~\eqref{eqS:H1} 
according to Eqs.~\eqref{eqS:scalar product in t} and \eqref{eqS:Floquet Hamiltonian elements}. We obtain
\newcommand{\mytau}[2]{\tau_\mathcal{#1Q}^{#2}}
\newcommand{\mytaud}[2]{\tau_\mathcal{Q#1}^{#2}}
\begin{equation}
\left(
\begin{tabular}{c|cccccccccccc}
$\cdots$ &$\mathcal{P}$0&$\mathcal{Q}$0&$\mathcal{R}$0
&$\mathcal{P}$1&$\mathcal{Q}$1&$\mathcal{R}$1&$\mathcal{P}2$&$\mathcal{Q}2$&$\mathcal{R}2$\\
\hline
$\mathcal{P}$0 & $\epsilon_{\mathcal{P}0}$ & $\mytau{P}{}$ & & $eV/2$ \\
$\mathcal{Q}$0 & $\mytaud{P}{}$ & $\epsilon_{\mathcal{Q}0}$ & $\mytaud{R}{}$ \\
$\mathcal{R}$0 & & $\mytau{R}{}$ & $\epsilon_{\mathcal{R}0}$&&&$eV/2$\\
$\mathcal{P}$1 & $eV/2$&&& $\epsilon_{\mathcal{P}1}$ & $\mytau{P}{}$ &&  $eV/2$  \\
$\mathcal{Q}$1 & &&& $\mytaud{P}{}$ & $\epsilon_{\mathcal{Q}1} $ & $\mytaud{R}{}$ \\
$\mathcal{R}$1 & &&$eV/2$ & & $\mytau{R}{}$ & $\epsilon_{\mathcal{R}1}$&&&$eV/2$ \\
$\mathcal{P}$2 & &&& $eV/2$ &&&$\epsilon_{\mathcal{P}2}$ & $\mytau{P}{}$ \\  
$\mathcal{Q}$2 & &&& &&& $\mytaud{P}{}$&$\epsilon_{\mathcal{Q}2}$&$\mytaud{R}{}$\\  
$\mathcal{R}$2 & && & && $eV/2$ &&$\mytau{R}{}$&$\epsilon_{\mathcal{R}2}$\\  
\end{tabular}
\right),
\label{eqS:HF1}
\end{equation}
where the matrix indexes are indicated by the row and column labels. For space reasons we introduced a short hand notation for the energies 
\begin{equation}
\epsilon_{\mathcal{\kappa}n}=\epsilon_\mathcal{\kappa}+n \hbar \omega.
\label{eqS:energies 1}
\end{equation} 
What we give in Eq.~\eqref{eqS:HF1} is a finite block of an infinite matrix, which is symbolized by the three dots in the left upper corner. The index $n$ takes negative as well as positive integer values, so that the first state which is not shown to the left in the first row would have index $\mathcal{R}(-1)$, while the next one continuing to the right at the row end would be $\mathcal{P}3$, and so on. Even though the matrix is infinite, the calculations are tractable because it has a periodic structure visible in Eq.~\eqref{eqS:HF1}, and following from Eq.~\eqref{eqS:Floquet Hamiltonian elements},
\begin{equation}
\langle \kappa (n+m) | H_F |\lambda (n^\prime+m) \rangle = \langle \kappa n | m \hbar \omega + H_F |\lambda n^\prime \rangle.
\label{eqS:periodic structure}
\end{equation}
Namely, upon a shift of the photon index the matrix elements are identical up to adding a constant on the diagonal.

Let us now suppose that the dots are tuned close to a single photon initial-final state resonance, corresponding to $N=1$ in the notation of Eq.~\eqref{eq:tac final general}. Denoting the corresponding energy explicitely as $\epsilon$, we have
\begin{equation}
\epsilon=\epsilon_\mathcal{R}\approx \epsilon_\mathcal{P} +\hbar \omega.
\label{eqS:resonance condition} 
\end{equation}
We are interested in the dynamics of the system starting in the initial state $\mathcal{P}$. Associating it with $\mathcal{P}1$ in the matrix $H_F$ [the choice of the value of the photon index is arbitrary, because of the periodic structure given in Eq.~\eqref{eqS:periodic structure}], we note that this state is, by Eq.~\eqref{eqS:resonance condition}, degenerate with state $\mathcal{R}0$. Together they span a degenerate subspace which we denote by projector 
\begin{equation}
P=|\mathcal{P}1\rangle \langle \mathcal{P}1| + |\mathcal{R}0\rangle \langle \mathcal{R}0|.
\end{equation}
If the matrix elements of $H_F$ between a state from subspace $P$ and another one from its complement $Q=1-P$ are much smaller than the difference of the corresponding diagonal entries, the dynamics produced by $H_F$ will be well approximated by restricting the basis to the degenerate subspace and taking the effects of other states perturbatively. We calculate the matrix elements of the effective Hamiltonian $H_P$ for the subspace $P$ using the following formula derived by the Brillouin-Wigner perturbation method \cite{ratner1990:JPC} 
\begin{equation}
H_P = PH_FP+\frac{PH_F^o Q}{E-H_F^d} \left( \sum_{p=0}^\infty \Big(\frac{Q H_F^o Q }{E-H_F^d} \Big)^p \right) Q H_F^oP.
\label{eqS:BW formula}
\end{equation}
The matrix divisions should be understood as $\frac{X}{Y}=X\cdot Y^{-1}$, and $H_F^{d/o}$ is the diagonal/off-diagonal part of the matrix $H_F$. The energy $E$ is the eigenvalue of the eigenstate of $H_P$, by which the equation is a self-consistent one (non-linear) in principle. However, this drawback is in practice not substantial, as one can solve order by order in the off-diagonal elements of $H_F$. The consecutively higher orders are indexed by the summation index $p$. To derive the results of the main text, we need only the lowest order of this formula, $p=0$ (the second order in off-diagonal matrix elements) for which one can replace $E$ by $\epsilon$ and get the standard linear Schroedinger equation with the Hamiltonian
\begin{equation}
H_P \approx PH_FP+ PH_F^o Q\frac{1}{\epsilon-H_F^d}  Q H_F^oP.
\label{eqS:effective Hamiltonian HP}
\end{equation}
The formula is valid if the off-diagonal elements and the energy differences of states within the subspace $P$ are much smaller than the energy denominator, which is required, respectively, for the convergence of the sum over $p$ in Eq.~\eqref{eqS:BW formula}, and for the replacement $E\to\epsilon$. Importantly, Eq.~\eqref{eqS:effective Hamiltonian HP} is valid for any dimension of $P$ and $Q$ and not only for our specific example of three states.
In practical calculations, it is usually straightforward to identify the most relevant virtual states that are to be retained in the subspace $Q$, while the majority of states can be neglected. Some states strictly do not contribute as there is no non-zero matrix element connecting them to the subspace $P$ (such as states with different number of electrons, or different spin).
The contribution of other states is subdominant in requiring more dot-dot hoppings (this is the case of the superexchange discussed in Sec.~V) or is negligible due to a large energy cost [the denominator in Eq.~\eqref{eqS:effective Hamiltonian HP}], 
which would be the case, {\it e.g.}, for states involving excited single particle orbitals (see Sec.~\ref{sec:errors} for an estimate).

We now return to our example, with $H_F$ given in Eq.~\eqref{eqS:HF1}. Trying to apply Eq.~\eqref{eqS:effective Hamiltonian HP}, however, we find that there is no term contributing to the matrix element $\langle \mathcal{P}1| H_P |\mathcal{R}0\rangle$ in the second order of $H_F^o$. Namely, to go from initial to the final state, it is needed to nearest-neighbor tunnel twice and absorb photon(s) once. Each of these corresponds to an off diagonal element in $H_F$, at minimum three off-diagonal terms together.
Using Eq.~\eqref{eqS:BW formula} in the next order, $p=1$, we get
\begin{equation}
\langle \mathcal{P}1| H_P |\mathcal{R}0\rangle \approx \frac{eV}{2\hbar \omega}\frac{\tau_{\mathcal{P}\mathcal{Q}} \,\tau_{\mathcal{Q}\mathcal{R}}}{\epsilon-\epsilon_{\mathcal{Q}0}}-\frac{eV}{2\hbar \omega}\frac{\tau_{\mathcal{P}\mathcal{Q}}\, \tau_{\mathcal{Q}\mathcal{R}}}{\epsilon-\epsilon_{\mathcal{Q}1}}.
\label{eqS:tac 0}
\end{equation}
This result shows that, apart from generating contributions in higher orders only, the matrix in Eq.~\eqref{eqS:HF1} potentially breaks the assumption of the off-diagonal elements being small compared to the energy differences. Namely, even though we always assume that neighboring dots are gated such that the tunneling amplitude is small compared to the detuning of the nearest empty state where an electron can hop in, the appearance of the ratio $eV/\hbar \omega$ would restrict the validity of our results to weak driving only, $eV \ll \hbar \omega$. To account for both of these issues, we introduce a unitary transformation as the last step necessary to obtain the cotunneling amplitudes given in the main text. 

We introduce a new basis, with states denoted by a tilde, by the following formula (this step goes beyond Ref.~\onlinecite{shirley1965:PR})
\begin{equation}
|\tilde{\mathcal{\kappa}n}\rangle = \sum_m J_{m}\left(\frac{eV_\kappa}{\hbar\omega}\right) |\kappa(n-m)\rangle,
\label{eqS:tilde state}
\end{equation} 
with $V_\mathcal{\kappa}$ the driving amplitude of the particular state, and $J_m(x)$ the Bessel function of the first kind. Using the sum rule for the Bessel functions [Eq.~\eqref{eqS:Bessel 1} and Eq.~\eqref{eqS:Bessel 3}], one can check that the new basis is also orthonormal and the matrix relating the old and new basis is, therefore, unitary with matrix elements
\begin{equation}
\langle \lambda m | \tilde{\kappa n} \rangle = \delta_{\kappa \lambda} J_{n-m} (eV_\mathcal{\kappa}/\hbar\omega).
\label{eqS:U}
\end{equation}
Using the time dependent representation of $|n\rangle$,
\begin{equation}
|\tilde{\kappa n}\rangle \to \sum_m J_{m}\Big(\frac{eV_\kappa}{\hbar\omega} \Big) e^{i (n-m) \omega t} |\kappa\rangle = e^{i n \omega t}|\tilde{\kappa} (t)\rangle,
\end{equation} 
along with the notation
\begin{equation}
|\tilde{\kappa}(t)\rangle \equiv |\kappa \rangle \exp\Big\{-\frac{i}{\hbar}\int_0^t {\rm d} t^\prime eV_\kappa \cos(\omega t^\prime)  \Big\},
\label{eqS:tdp ket}
\end{equation}
one can understand the choice in Eq.~\eqref{eqS:tilde state} as accommodating the basis to include the accumulated phase from the oscillating part of the energy. 

In this basis the matrix $H_F$ takes the form
\renewcommand{\mytau}[2]{\tau_\mathcal{#1Q}^{(#2)}}
\renewcommand{\mytaud}[2]{\tau_\mathcal{Q#1}^{(#2)}}
\begin{equation}
\left(
\begin{tabular}{c|ccccccccc}
$\cdots$ &$\tilde{\mathcal{P}0}$&$\tilde{\mathcal{Q}0}$
&$\tilde{\mathcal{R}0}$&
$\tilde{\mathcal{P}1}$&$\tilde{\mathcal{Q}1}$&$\tilde{\mathcal{R}1}$
&$\tilde{\mathcal{P}2}$&$\tilde{\mathcal{Q}2}$&$\tilde{\mathcal{R}2}$ \\
\hline
$\tilde{\mathcal{P}0}$ & $\epsilon_{\mathcal{P}0}$ & $\mytau{P}{0}$ & & & $\mytau{P}{-1}$ &&& $\mytau{P}{-2}$  \\
$\tilde{\mathcal{Q}0}$& $\mytaud{P}{0}$ & $\epsilon_{\mathcal{Q}0}$ & $\mytaud{R}{0}$ &$\mytaud{P}{1}$&& $\mytaud{R}{1}$ &$\mytaud{P}{2}$&& $\mytaud{R}{2}$  \\
$\tilde{\mathcal{R}0}$  & & $\mytau{R}{0}$ & $\epsilon_{\mathcal{R}0}$ & & $\mytau{R}{-1}$&&&$\mytau{R}{-2}$\\
$\tilde{\mathcal{P}1}$ & &$\mytau{P}{1}$&& $\epsilon_{\mathcal{P}1}$ & $\mytau{P}{0}$ && & $\mytau{P}{-1}$ \\
$\tilde{\mathcal{Q}1}$& $\mytaud{P}{-1}$& &$\mytaud{R}{-1}$  & $\mytaud{P}{0}$ & $\epsilon_{\mathcal{Q}1}$ & $\mytaud{R}{0}$ & $\mytaud{P}{1}$& &$\mytaud{R}{1}$  \\
$\tilde{\mathcal{R}1}$ &  & $\mytau{R}{1}$ && & $\mytau{R}{0}$ & $\epsilon_{\mathcal{R}1}$&&$\mytau{R}{-1}$\\
$\tilde{\mathcal{P}2}$ & &$\mytau{P}{2}$&& &$\mytau{P}{1}$&&$\epsilon_{\mathcal{P}2}$&$\mytau{P}{0}$&\\  
$\tilde{\mathcal{Q}2}$& $\mytaud{P}{-2}$& &$\mytaud{R}{-2}$  & $\mytaud{P}{-1}$ & & $\mytaud{R}{-1}$ & $\mytaud{P}{0}$& $\epsilon_{\mathcal{Q}2}$ &$\mytaud{R}{0}$  \\
$\tilde{\mathcal{R}2}$ & &$\mytau{R}{2}$&& &$\mytau{R}{1}$&&&$\mytau{R}{0}$&$\epsilon_{\mathcal{R}2}$
\end{tabular}
\right).
\label{eqS:HF2}
\end{equation}
The terms $eV/2$ were removed from the off-diagonal on the expense of generating more tunneling elements. Unlike in the original basis, where a time independent operator is diagonal in the photon index, the inter-dot tunneling Hamiltonian now has elements also between states with different integer indexes
\begin{equation}
\tau_{\kappa\lambda}^{(n-m)} = \langle \tilde{\kappa n} | H_T| \tilde{\lambda m} \rangle =  \tau_{\kappa \lambda} J_{n-m}(eV_{\kappa\lambda}/\hbar\omega),
\label{eqS:tau index definition}
\end{equation} 
with $V_{\kappa\lambda} =V_{\kappa}-V_{\lambda}$ the voltage amplitude drop between the two states. Since the Bessel functions are not larger than one for any real parameter, the matrix $H_F$ in the newly adopted basis is suitable for perturbative calculations even for a strong driving, $eV \gg \hbar \omega$. 

We note that such complete removal of the driving terms proportional to the voltage is possible because of the form of the driving part of the Hamiltonian that we chose in Eq.~\eqref{eq:HD}. By that choice we neglect the spatial deformation of dot states induced by the electric field. This simplification can make a qualitative difference only if such terms would break some symmetry which otherwise blocks tunnelings.\cite{platero2004:PR} There is no such symmetry in our case. Also, this simplification is not essential for using the Shirley technique.

The propagator in the transformed basis is 
\begin{equation}
\begin{split}
U(t,t_0)=
\sum_{\kappa\lambda n} 
|\tilde{\kappa}(t)\rangle \langle \tilde{\lambda}(t_0)| 
e^{i n \omega t} \langle \tilde{\kappa n} | U_F |\tilde{\lambda 0} \rangle,
\end{split}
\label{eqS:propagator new basis}
\end{equation}
where the bra vector $\langle \tilde{\kappa}(t)|$ is defined by a complex conjugation of Eq.~\eqref{eqS:tdp ket}, and $U_F=\exp\{-(i/\hbar)H_F (t-t_0) \}$. Equation \eqref{eqS:propagator new basis} is a complete analogue to Eq.~\eqref{eqS:Shirley propagator}.

We now evaluate off-diagonal elements of the effective Hamiltonian in the transformed basis. Using Eq.~\eqref{eqS:HF2} in Eq.~\eqref{eqS:effective Hamiltonian HP} we get 
\begin{equation}
\langle \tilde{\mathcal{P}1}| H_P |\tilde{\mathcal{R}0} \rangle =\sum_{n=-\infty}^\infty \frac{\mytau{P}{1-n}\mytaud{R}{n}}{\epsilon-\epsilon_{\mathcal{Q}} - n \hbar \omega}.
\label{eqS:tac 1}
\end{equation}
The validity of the formula follows from conditions on the validity of Eq.~\eqref{eqS:effective Hamiltonian HP}, which were stated therein. In terms of the parameters used here, both the photon assisted tunneling amplitudes and the degeneracy detuning should be smaller than the energy denominator
\begin{equation}
|\mytau{P}{1-n}|, |\mytaud{R}{n}|, |\epsilon_\mathcal{R} - \epsilon_\mathcal{P} -\hbar \omega| \ll |\epsilon-\epsilon_\mathcal{Q} - n\hbar \omega|,
\label{eqS:condition on validity}
\end{equation}
for each term in the summation over the photon index $n$.

Equation \eqref{eqS:tac 1} takes the microwave field into account to all orders, enumerated by index $n$. It would be difficult to include such higher order processes using Eq.~\eqref{eqS:HF1}, as they correspond to higher order of perturbation expansion in Eq.~\eqref{eqS:BW formula}. The correspondence between the two bases can be established upon expanding the Bessel functions in their argument. For weak driving the tunneling amplitudes $\tau^{(n)}$ fall off exponentially with $|n|$. In such a case, retaining only the leading order terms, $n=0,1$, using Eq.~\eqref{eqS:tau index definition}, expanding the Bessel functions up to the lowest order using Eq.~\eqref{eqS:Bessel 4}, Eq.~\eqref{eqS:tac 1} reduces to Eq.~\eqref{eqS:tac 0}.

We can now easily generalize to $N$ photon resonance, and to more intermediate states $\mathcal{Q}$. The former means that the initial and final state energies differ by $N \hbar \omega$, and the generalization amounts to replacing the upper index 1 by $N$ in Eq.~\eqref{eqS:tac 1}. The latter means a summation over the intermediate states, as contributions from subspace $Q$ are additive in Eq.~\eqref{eqS:effective Hamiltonian HP}. With these generalizations we get
\begin{equation}
\langle \tilde{\mathcal{P}N}| H_P |\tilde{\mathcal{R}0} \rangle = \sum_{\mathcal{Q}} \sum_{n=-\infty}^\infty \frac{\mytau{P}{N-n}\mytaud{R}{n}}{\epsilon-\epsilon_{\mathcal{Q}} - n \hbar \omega}.
\label{eqS:tac final}
\end{equation}
Denoting the left hand side as $\tau_{co}$ and relating to the parameters of the original Hamiltonian by Eqs.~\eqref{eqS:tau definition}, \eqref{eqS:resonance condition}, and \eqref{eqS:tau index definition}, we obtain Eq.~\eqref{eq:tac final general}.

We remind that the ``initial-final'' state resonance discussed in the main text refers to a configuration in which the initial and final states differ in energy. The system can make the transition between these two states only if a non-zero number of photons is absorbed in total, with the photon number $N$ given by the resonance condition $N \hbar \omega$ equal to the initial and final state energy difference. In the often met case of the initial to final state detuning being much smaller than the virtual states offsets, so that one can neglect the photon energies in the denominator of Eq.~\eqref{eqS:tac final}, the sum over the photon index can be evaluated using Eq.~\eqref{eqS:Bessel 1} to arrive at
\begin{equation}
\langle \tilde{\mathcal{P}N}| H_P |\tilde{\mathcal{R}0} \rangle \approx \left( \sum_{\mathcal{Q}} \frac{\tau_{\mathcal{P}\mathcal{Q}} \tau_{\mathcal{Q}\mathcal{R}}}{\epsilon-\epsilon_{\mathcal{Q}}} \right) J_N\left( \frac{eV_{\mathcal{P}\mathcal{R}}}{\hbar \omega} \right),
\label{eqS:tac initial final simplified}
\end{equation}
Eq.~\eqref{eq:tac initial final simplified} of the main text. In this configuration the photon assisted cotunneling is proportional to a cotunneling without any driving [the first bracket in Eq.~\eqref{eqS:tac initial final simplified}] times a suppression factor depending on the relative amplitude of driving of the initial and final states. The role of microwaves is to allow for the observation of the cotunneling by coupling the resonant states $|\tilde{\mathcal{P}N}\rangle$ and $|\tilde{\mathcal{R}0}\rangle$. The result also shows that in this regime driving the mediating states is ineffective.

The ``virtual resonance'', on the other hand, corresponds to $N=0$, so that energy does not have to be provided by microwaves in order for the transition to occur. Rather, the resonance now means that the microwave frequency is such that a virtual state $\mathcal{Q}$ becomes quasi-degenerate with the initial and final state energies upon adding the energy of $n$ photons. Based on a much smaller energy denominator, the sum in Eq.~\eqref{eqS:tac final} can be approximated by this single term $n$ and state $\mathcal{Q}$ (or a few close-by states).

Once the cotunneling amplitude is calculated, the dynamics of the system can be found using the effective Hamiltonian $H_P$, which in the subspace $\{|\tilde{\mathcal{P}N}\rangle, |\tilde{\mathcal{R}0}\rangle \}$ takes the form
\begin{equation}
H_P=\left( 
\begin{tabular}{cc}
$\epsilon_\mathcal{P}+N \hbar \omega$ & $\tau_{co}$\\
 $\tau_{co}^\dagger$ & $\epsilon_\mathcal{R}$
\end{tabular} 
\right),
\end{equation}
and the propagator in the original Hilbert space basis $\{|\tilde{\mathcal{P}}\rangle,|\tilde{\mathcal{R}}\rangle\}$, with the phase factors defined in Eq.~\eqref{eqS:tdp ket}, takes the form 
\begin{equation}
U(t_0,t)=
\left( 
\begin{tabular}{cc}
$e^{i N \omega (t-t_0)}$\quad & 0 \\
 0 \quad & $1$ 
\end{tabular} 
\right) \exp\left\{ -\frac{i}{\hbar} H_P (t-t_0) \right\}.
\label{eqS:phases in propagator}
\end{equation}
Close to resonance $\epsilon_\mathcal{R}\approx \epsilon_\mathcal{P} + N \hbar \omega $ the system will display Rabi oscillations with frequency $\tau_{co}$. This finishes the derivation and interpretation of the photon assisted cotunneling amplitudes. 

\subsection*{Bessel functions properties}

Here we list a few properties of Bessel functions which are needed in the discussed derivations,
\begin{align}
&e^{iz\sin \phi} = \sum_{n=-\infty}^\infty  J_{n}(z) e^{i n \phi},
\label{eqS:Bessel 5}\\
&J_n(z_1\pm z_2) = \sum_{k=-\infty}^\infty  J_{n\mp k}(z_1) J_k(z_2),
\label{eqS:Bessel 1}\\
&J_{-n}(x)=(-1)^nJ_n(x),
\label{eqS:Bessel 2}\\
&J_n(0)=\delta_{n,0},
\label{eqS:Bessel 3}\\
&J_n(x)\approx(x/2)^{n}/n!.
\label{eqS:Bessel 4}
\end{align}
In the last equation, the result is given in the leading order expansion around $x=0$ assuming $n$ to be a positive integer.

\section{Effective Hamiltonians for the few level model: derivations of Eqs.~\eqref{eq:2e3d} and \eqref{eq:3e3d}}

In this section we introduce the few level model for which the effective Hamiltonians for a three dot $(A,B,C)$ structure, given in Eq.~\eqref{eq:2e3d} and Eqs.~\eqref{eq:3e3d}, are valid. We consider two particle states in the middle dot ($B$) which can be built out of the two lowest single particle orbitals ($i=1,2$). Denoting the corresponding fermionic operators as $c_{B1\sigma}$ and $c_{B2\sigma}$, these comprise the following six states
\begin{subequations}
\begin{eqnarray}
|S\rangle & = & 2^{-1/2} (c_{B1\uparrow}^\dagger c_{B2\downarrow}^\dagger -c_{B1\downarrow}^\dagger c_{B2\uparrow}^\dagger)  |0\rangle,\\
|S^\prime_1\rangle & = & c_{B1\uparrow}^\dagger c_{B1\downarrow}^\dagger |0\rangle,\\
|S^\prime_2\rangle & = & c_{B2\uparrow}^\dagger c_{B2\downarrow}^\dagger |0\rangle,\label{eqS:B1c}\\
|T_0\rangle & = & 2^{-1/2} (c_{B1\uparrow}^\dagger c_{B2\downarrow}^\dagger + c_{B1\downarrow}^\dagger c_{B2\uparrow}^\dagger)  |0\rangle,\\
|T_+\rangle & = & c_{B1\uparrow}^\dagger c_{B2\uparrow}^\dagger |0\rangle,\\
|T_-\rangle & = & c_{B1\downarrow}^\dagger c_{B2\downarrow}^\dagger |0\rangle,
\end{eqnarray}
\end{subequations}
where state $|0\rangle$ denotes an empty dot.

To get an analytically manageable model, we now restrict ourselves to a specific configuration. We assume the three dot system is electrostatically gated such that the middle dot single particle ground state is well below the aligned outer dot ones. The relevant virtual states then do not include doubly occupied outer dots and it is enough to consider only the lowest orbital level $i=1$ in these. To calculate the cotunneling amplitudes, we need the matrix elements of the tunneling Hamiltonian $H_T$, Eq.~\eqref{eq:HT}, between states of the three dot system. A short calculation gives the following auxiliary result
\begin{equation}
\begin{split}
H_T &c_{A1\sigma}^\dagger c_{Bis}^\dagger |0 0 0 \rangle = -\tau^{CB}_{1i} |\sigma 0 s\rangle -\tau^{AB}_{1 i} \delta_{\sigma\overline{s}} \sigma p_i |0 S^\prime_i 0 \rangle\\
&  + \tau^{AB}_{1\overline{i}} \left( \frac{\delta_{\sigma\overline{s}}}{\sqrt{2}} (\sigma |0 S 0\rangle - p_i |0 T_0 0\rangle ) - \delta_{\sigma s} p_i |0 T_\sigma 0\rangle\right),
\end{split}
\label{eqS:HT matrix element 0}
\end{equation}
where we use an overline to denote the complementary index, $\overline{1}=2$ and $\overline{2}=1$, and similarly for $\sigma=\uparrow, \downarrow$. If the spin index is used as a factor, it should be understood as $\sigma=+1$, and $-1$, corresponding to $\uparrow$, and $\downarrow$, respectively. Finally, we introduce additional signs by $p_1=1=-p_2$. 

The state on the left hand side of Eq.~\eqref{eqS:HT matrix element 0} is the initial state $|\sigma s 0\rangle$  for $i=1$. The action of the tunneling Hamiltonian on the final state of the form $|0 s \sigma\rangle$ can be obtained by swapping the labels $A$ and $C$. The right hand side shows which virtual states are connected by a single particle hopping to the initial and final states. We now neglect the contribution from the virtual states in the first line of Eq.~\eqref{eqS:HT matrix element 0} based on their higher energy offsets compared to the states retained, which are those in the second line. The same would result from an assumption that the inter-dot tunneling is dominated by tunneling into an excited state $|\tau_{11}|\ll |\tau_{12}|$. This is expected to be the case in tunnel coupled dots, because of a larger spatial extent of an excited state, and was exploit in the spin measurement by conversion to charge in the experiment of Ref~\onlinecite{hanson2005:PRL}.
 Finally, in the case of a virtual resonance, which we consider below, it is the microwave frequency which selects the relevant set of virtual states. We assume this is the set of the singlet $S$ and the three triplets $T_{0,\pm}$ in the middle dot, with all other states displaced by a large energy, on the scale of the single dot orbital excitation energy or the charging energy. With any of these reasonings, 
\begin{equation}
\begin{split}
H_T |\sigma s 0 \rangle &=  \tau_{12}^{AB} \left(\frac{\delta_{\sigma \overline{s}}}{\sqrt{2}} (\sigma |0 S 0\rangle-|0 T_0 0\rangle) -\delta_{\sigma s} |0 T_\sigma 0\rangle  \right),\\
H_T |0 s \sigma \rangle &=  \tau_{12}^{CB} \left(\frac{\delta_{\sigma \overline{s}}}{\sqrt{2}} (\sigma |0 S 0\rangle-|0 T_0 0\rangle) -\delta_{\sigma s} |0 T_\sigma 0\rangle  \right),
\end{split}
\label{eqS:HT matrix element}
\end{equation}
is a good approximation for the calculation of the cotunneling amplitudes.

With the three dots gated as described, the two electron ground state (the degenerate subspace $P$) comprises eight states, $\mathcal{P},\mathcal{R} \in \{ |\sigma s 0\rangle, |0 s^\prime \sigma^\prime\rangle \}$ with the spins $s,s^\prime,\sigma,\sigma^\prime \in \{ \uparrow,\downarrow\}$. Taking into account that these states are degenerate without the microwave assistance, we can apply Eq.~\eqref{eqS:tac final} with $N=0$:
\begin{equation}
\langle \tilde{\mathcal{P}0} | H_P | \tilde{\mathcal{R}0} \rangle=
{\sum_\mathcal{Q} }^\prime  \frac{\langle \mathcal{P} | H_T | \mathcal{Q} \rangle \langle \mathcal{Q} | H_T | \mathcal{R} \rangle}{\epsilon-\epsilon_{\mathcal{Q}}+n \hbar \omega}
 J^2_n\Big(\frac{eV}{\hbar\omega} \Big).
\label{eqS:HP2 0}
\end{equation}
The prime restricts the summation to states $\mathcal{Q}=|0k0\rangle$ with $k \in \{ S, T_0,T_\pm \}$, as follows from assumption of the microwave frequency being tuned close to these four states for some integer multiple $n$ of the photon energy. We also assumed the middle dot is driven at amplitude $V$, which gives $V_{\mathcal{P}\mathcal{Q}}=V_{\mathcal{R}\mathcal{Q}}=-V$. For simplicity, we also consider a symmetric structure with phases of single electron states chosen such that $\tau_{12}^{AB} = \tau_{12}^{CB}$.
With this we evaluate the effective Hamiltonian $H_P$ using Eqs.~\eqref{eqS:HT matrix element} and \eqref{eqS:HP2 0}. Due to the spin conservation, it is block diagonal. For the unpolarized subspace we get
\begin{equation}
\left(
\begin{tabular}{c|cccc}
& $|\uparrow \downarrow 0\rangle$ &$|\downarrow \uparrow  0\rangle$ &$|0 \uparrow \downarrow\rangle$ &$| 0 \downarrow \uparrow  \rangle$ \\
\hline
$|\uparrow \downarrow 0\rangle$ & $\epsilon_+$ & $\epsilon_-$ & $\epsilon_-$ & $\epsilon_+$\\
$|\downarrow \uparrow  0\rangle$ & $\epsilon_-$ & $\epsilon_+$ & $\epsilon_+$ & $\epsilon_-$\\
$|0 \uparrow \downarrow\rangle$ & $\epsilon_-$ & $\epsilon_+$ & $\epsilon_+$ & $\epsilon_-$\\
$| 0 \downarrow \uparrow  \rangle$ & $\epsilon_+$ & $\epsilon_-$ & $\epsilon_-$ & $\epsilon_+$
\end{tabular}
\right),
\label{eqS:HP2 1}
\end{equation}
where the energies $\epsilon_\pm $ are defined as
\begin{equation}
\epsilon_\pm = \tau_V^2 \frac{ \delta_{T_0}^{-1} \pm  \delta_S^{-1} }{2}.
\label{eqS:epm}
\end{equation}
The energy offsets are defined as
\begin{equation}
\delta_\mathcal{Q}={\epsilon_\mathcal{P}-\epsilon_{\mathcal{Q}} + n \hbar \omega},
\end{equation}
and refer to the energy of the states in the degenerate subspace $\epsilon_\mathcal{P}$. The microwave assisted tunneling amplitude is given by
\begin{equation}
\tau_V^2 = J^2_n\Big(\frac{eV}{\hbar\omega} \Big) \tau_{12}^{AB} \tau_{21}^{BC}.
\label{eqS:tauV}
\end{equation}
To simplify the notation, we omitted the tilde and the photon index from the basis states in Eq.~\eqref{eqS:HP2 1}. Namely, all photon indexes of states $\mathcal{P}$ and $\mathcal{R}$ are identical for the case of virtual resonance ($N=0$) and can be chosen, {\it e.g.}, to be zero while there are no additional phases arising in the propagator apart from those generated by $H_P$ as follows from Eq.~\eqref{eqS:phases in propagator}. 
Using analogous notation in further, we get for the polarized subspaces
\begin{equation}
\left(
\begin{tabular}{c|cc}
& $|\sigma \sigma 0\rangle$ &$|0 \sigma \sigma \rangle$\\
\hline
$|\sigma \sigma 0\rangle$ & $\epsilon_\sigma$ & $\epsilon_\sigma$\\
$|0 \sigma \sigma  \rangle$ & $\epsilon_\sigma$ & $\epsilon_\sigma$ \\
\end{tabular}
\right),
\label{eqS:HP2 2}
\end{equation}
with $\epsilon_\sigma=\tau_V^2 / \delta_{T_\sigma}$. Equations \eqref{eqS:HP2 1} and \eqref{eqS:HP2 2} are the result given in Eq.~\eqref{eq:2e3d}, as we now show by introducing the following notations. Instead of the tensor product basis, we define an outer and inner spins by introducing Pauli matrices
\begin{subequations}
\begin{eqnarray}
\boldsymbol{\sigma}^o &=& \mathds{1}_A \otimes \mathds{1}_B \otimes \boldsymbol{\sigma}_C +\boldsymbol{\sigma}_A \otimes \mathds{1}_B \otimes \mathds{1}_C,\\
\boldsymbol{\sigma}^i &=& \mathds{1}_A \otimes \boldsymbol{\sigma}_B  \otimes \mathds{1}_C,
\end{eqnarray}
\end{subequations}
with a sigma matrix acting on an empty state of a dot defined to result in zero. Another pseudo-spin degree of freedom relates to the position of the charge in the outer dots through operators
\begin{subequations}
\begin{eqnarray}
\eta_x &=&\sum_{s\sigma} |\sigma s 0\rangle \langle 0 s \sigma | + |0 s \sigma \rangle \langle \sigma s 0|,\\
\eta_y &=&\sum_{s\sigma} i |0 s \sigma \rangle \langle \sigma s 0| - i|\sigma s 0\rangle \langle 0 s \sigma | ,\\
\eta_z &=&\sum_{s\sigma} |\sigma s 0\rangle \langle \sigma s 0 | - |0 s \sigma \rangle \langle 0 s \sigma |.
\end{eqnarray}
\end{subequations}
With these operators, the effective Hamiltonian in the unpolarized subspace, Eq.~\eqref{eqS:HP2 1}, is
\begin{equation}
P_0 H_P P_0 = P_0 \Big[(1+ \eta_x) \Big(\epsilon_+ + \frac{\sigma^o_+ \sigma^i_- + \sigma^o_- \sigma^i_+}{4} \epsilon_-\Big)\Big] P_0, 
\label{eqS:HP2 3}
\end{equation}
and in the polarized subspace, Eq.~\eqref{eqS:HP2 2}, is
\begin{equation}
P_\sigma H_P P_\sigma = P_\sigma (1+ \eta_x) \epsilon_\sigma P_\sigma.
\label{eqS:HP2 4}
\end{equation}
We defined $P_s$ as a projector to the subspace with a definite spin z-projection $s=0,\pm1$ by
\begin{subequations}
\begin{eqnarray}
P_1 &=&(1/4) (1+\sigma^o_z) (1+\sigma^i_z),\\
P_{-1} &=&(1/4) (1-\sigma^o_z) (1-\sigma^i_z),\\
P_0&=&1-P_1-P_{-1}.
\end{eqnarray}
\end{subequations}
With the help of these, adding Eq.~\eqref{eqS:HP2 3} and \eqref{eqS:HP2 4} finishes the way to Eq.~\eqref{eq:2e3d}.

Let us now turn to the configuration of each dot singly occupied (the exchange configuration). The ground state manifold comprises eight degenerate states, which we denote by 
\begin{equation}
|\sigma \pi \rho \rangle = c_{A1\sigma}^\dagger  c_{B1\pi}^\dagger c_{C1\rho}^\dagger | 0 0 0 \rangle.
\end{equation}
To derive the effective Hamiltonian for this subspace, we proceed analogously to the previous case. Under the assumptions that the electrostatic and microwave tuning selects the four doubly occupied states of the middle dot as the only relevant excited states, the cotunneling amplitudes are given by Eq.~\eqref{eqS:HP2 0}. 
In this case the matrix elements of the tunneling Hamiltonian can be written as
\begin{equation}
H_T |\sigma \pi \rho \rangle = c_{A1\sigma}^\dagger H_T |0 \pi \rho \rangle +  c_{C1\rho}^\dagger H_T |\sigma \pi 0 \rangle.
\label{eqS:HT matrix element 2}
\end{equation}
where the terms on the right hand side are given by Eq.~\eqref{eqS:HT matrix element}. A short calculation gives
\begin{equation}
\left(
\begin{tabular}{c|ccc}
& $|\uparrow \uparrow \downarrow \rangle$ &$|\uparrow \downarrow \uparrow  \rangle$ &$| \downarrow \uparrow \uparrow \rangle$\\
\hline
$|\uparrow \uparrow \downarrow \rangle$ & $\epsilon_+ + \epsilon_\uparrow$ & $\epsilon_-$ & 0\\
$|\uparrow \downarrow \uparrow  \rangle$ & $\epsilon_-$ & $2\epsilon_+$ & $\epsilon_-$\\
$|\downarrow \uparrow \uparrow \rangle$ & 0 & $\epsilon_-$ & $\epsilon_+ + \epsilon_\uparrow$
\end{tabular}
\right),
\label{eqS:HP3 1}
\end{equation}
and 
\begin{equation}
\left(
\begin{tabular}{c|ccc}
& $|\downarrow \downarrow \uparrow \rangle$ &$|\downarrow \uparrow \downarrow  \rangle$ &$| \uparrow \downarrow \downarrow \rangle$\\
\hline
$|\downarrow \downarrow \uparrow \rangle$ & $\epsilon_+ + \epsilon_\downarrow$ & $\epsilon_-$ & 0\\
$|\downarrow \uparrow \downarrow  \rangle$ & $\epsilon_-$ & $2\epsilon_+$ & $\epsilon_-$\\
$|\uparrow \downarrow \downarrow \rangle$ & 0 & $\epsilon_-$ & $\epsilon_+ + \epsilon_\downarrow$
\end{tabular}
\right),
\label{eqS:HP3 2}
\end{equation}
for the partially polarized subspaces with the total spin $s=\pm 1/2$, and 
\begin{equation}
\langle \sigma \sigma \sigma | H_P | \sigma \sigma \sigma \rangle = 2 \epsilon_\sigma,
\label{eqS:HP3 3}
\end{equation}
for a fully polarized subspace with the total spin $s=\pm 3/2$. Here, we again used the energy notations defined in Eq.~\eqref{eqS:epm} and below Eq.~\eqref{eqS:HP2 2}.

These equations give the effective Hamiltonian stated in the main text by identifying the diagonal elements in Eqs.~\eqref{eqS:HP3 1}--\eqref{eqS:HP3 3} as $\mathcal{H}_d+\mathcal{H}_z$ and the off-diagonal elements as the term $\mathcal{H}_{SWAP}$. 
 
\section{Long distance amplitudes: derivation of Eqs.~\eqref{eq:tac long} and \eqref{eq:leakage}}

Here we derive the long-distance scaling of the PACT amplitudes. To this end, we expand the dot model discussed previously from three to $M-1$ dots indexed by an integer index $j=1,\ldots,M-1$ and investigate cotunneling amplitudes for electron transport and spin exchange between two selected dots. These are labeled as $A$, $C$ and placed within the array at positions $j_A$ and $j_C$ with the distance $d = j_C-j_A$ between them. We are interested in the amplitude behavior for $d \gg 1$, which is what we mean by the long-distance cotunneling: a coupling by coherently leapfrogging many dots at once.

We assume the structure is gated such that (singly occupied) dots throughout the array are aligned in energy. These aligned states should be the subspace of singlet $S$ and triplet $T_0$ tuned to degeneracy by, {\it e.g.}, external magnetic field, or be it some higher excited states with negligible exchange splitting, or by forming a double dot with an auxiliary dot in a nearby array, in which case the exchange can be easily modified by a detuning bias.\cite{petta2005:S, hanson2007:RMP, brunner2011:PRL, stepanenko2012:PRB} The condition on the degeneracy is that the splitting should be much smaller than the tunneling matrix element between these states in neighboring dots (which grows for higher excited states because of an increased wave function spatial extent). For identical dots such alignment corresponds to alignment of the dots' ground states as depicted on Fig.~1b. The two manipulated dots are detuned from the others in the array. Let us first consider a situation where 
dots A and C contain a single 
electron in total, and we intend to induce non-local electron transfer between them. It is then desirable to tune them such that there is a single electron excited state aligned with the band. Also, both these dots should be driven at the same frequency (its value specified below). This configuration is depicted in Fig.~1c.

To keep the problem analytically tractable, we approximate the virtual band by a uniformly coupled linear chain of $M-1$ sites. That is, we neglect the heterogeneity of the array induced by detuning dots A and C and by their driving (the latter inessential at weak driving) and assume that the matrix elements of $H_T$ between the aligned states do not depend on the position in the array, $j$, and denote them by $\tau$, with the scale given by the tunneling matrix element between nearest neighbor single particle states ($\tau^{AB}_{22}$ in the notation of the previous sections).

The states of the virtual band are then described by wavefunctions
\begin{equation}
|\Psi_q\rangle=\sum_{\sigma \alpha} \Psi_q(j_\alpha) c_{\alpha 2 \sigma}^\dagger c_{A 1 \sigma} |G\rangle,
\end{equation}
with $|G\rangle$ representing the ground state of the array, from which the electron from the dot $A$ is virtually excited into the band. Since the excited electron can be on any dot, the index $q$ takes values from $1$ to $M-1$. For uniform hoppings, these wave functions are those of a linear chain with zero boundary conditions
\begin{equation}
\Psi_q(j)=\sqrt{\frac{2}{M}} \sin \left( \frac{\pi}{M}q j \right).
\label{eqS:wave functions}
\end{equation}
The corresponding energies are
\begin{equation}
\epsilon_q=\epsilon_B+4\tau \sin^2 \left( \frac{\pi q}{2M} \right),
\label{eqS:energies}
\end{equation}
where by $\epsilon_B$ we denote the bottom of the band.

With the above definitions, the cotunneling amplitude follows from Eq.~\eqref{eqS:HP2 0} as
\begin{equation}
\tau_{co} = \sum_{q=1}^{M-1} \frac{\tau_{Aq}^{(n)}\tau_{qC}^{(-n)}}{\delta-\epsilon_q},
\label{eqS:long discrete}
\end{equation}
where we assumed both dots $A$ and $C$ are driven at the same voltage amplitude $V$ and introduced the microwave tunable offset from the band bottom defined as
\begin{equation}
\delta=\epsilon_A-\epsilon_B + n \hbar \omega.
\end{equation}
The tunneling amplitude from dot $\alpha$ into the delocalized band is
\begin{equation}
\tau_{\alpha q}=\tau_{AB}^{12} \Big( \Psi_q (j_\alpha-1)+ \Psi_q (j_\alpha+1) \Big),
\label{eqS:tunneling me aux}
\end{equation}
as it can occur through any of the two nearest neighbors of the dot $\alpha$.

In this section, we are interested in the possible maximum of the cotunneling and in boosting it by diminishing the denominator in Eq.~\eqref{eqS:long discrete} by the microwave matching. However, Eq.~\eqref{eqS:long discrete} formally diverges upon tuning its denominator to zero. This is unphysical as it violates the condition in Eq.~\eqref{eqS:condition on validity}. A simple and physically appealing regularization of such spurious divergences is to consider explicitly the  decoherence (energy smearing) by inserting an imaginary part into the virtual state energies, $\epsilon_\mathcal{Q} \to \epsilon_\mathcal{Q} -i\gamma$. The scale for $\gamma$ is set by the scale of the coherence or life time of the dot states themselves, which we take of order of ten nanoseconds corresponding to $\gamma \sim 50$ neV (a typical scale for energy of random nuclear magnetic fields,\cite{erlingsson2001:PRB,khaetskii2002:PRL} and also phonon induced orbital relaxation\cite{khaetskii2000:PRB, stano2006:PRB}).

With this change of the cotunneling formula, the diagonal elements of the effective Hamiltonian $H_P$ become complex. Let us consider the imaginary part of such an element,
\begin{equation}
- {\rm Im} [ \langle \tilde{\mathcal{P}0}| H_P |\tilde{\mathcal{P}0} \rangle ] = \pi \sum_{q=1}^{M-1} |\tau_{Aq}^{(n)}|^2  g_\gamma(\delta-\epsilon_q),
\label{eqS:leakage 1}
\end{equation}
where we denoted
\begin{equation}
g_\gamma(\delta-\epsilon_q) = \frac{1}{\pi}\frac{\gamma}{(\delta-\epsilon_q)^2 + \gamma^2}.
\label{eqS:lorentz}
\end{equation}
We recognize in Eq.~\eqref{eqS:leakage 1} the Fermi's golden rule formula
\begin{equation}
\hbar \Gamma = 2\pi \int {\rm d}\epsilon \sum_q |\langle \tilde{\mathcal{P}0}| H_T |\tilde{\mathcal{Q}n} \rangle|^2 \delta(\epsilon_\mathcal{P}-\epsilon) g_\gamma(\epsilon-\epsilon_{\mathcal{Q}n}),
\label{eqS:FGR}
\end{equation}
for the decay rate of state $|\tilde{\mathcal{P}0} \rangle$ due to coupling to the band of states if these states have the energies given as the Lorentz probability distribution,  Eq.~\eqref{eqS:lorentz}. The factor of 2 is a conversion between a Hamiltonian matrix element and the rate for the change of the probability. We, therefore, define
\begin{equation}
\hbar \Gamma = - 2\, {\rm Im} [ \langle \tilde{\mathcal{P}0}| H_P |\tilde{\mathcal{P}0} \rangle ] ,
\label{eqS:leakage}
\end{equation}
as the leakage rate by which the initial state leaves the computational subspace. The cotunneling is to be compared to this rate and the leakage must be much smaller than the cotunneling for any useful manipulations.

Now we are ready to evaluate the relevant rates. We are interested in the limit of a long array, $M\gg 1$, and a large distance between the dots, $d\gg 1$. This allows us to adopt several approximations, which make it possible to obtain results in an explicit analytical form. First, we neglect the difference of the wave functions arguments in Eq.~\eqref{eqS:tunneling me aux}, by putting $\tau_{q \alpha} \approx 2 \tau \Psi_q (j_\alpha)$. Second, we assume the dots $A$ and $C$ are not too close to the array boundary (we will quantify this condition later), so we can change the boundary conditions to different ones, {\it e.g.} periodic, and replace the wave functions in Eq.~\eqref{eqS:wave functions} by plane waves
\begin{equation}
\Psi_q(j) \simeq \sqrt{\frac{1}{M}} \exp \left( \frac{\pi}{M}q j \right).
\label{eqS:plane waves}
\end{equation}
Finally, assuming the regime in which the leakage is much smaller than the cotunneling, we approximate the latter energy denominator by its real part only. Using the notation of Eq.~\eqref{eqS:tauV} for the overall scale, we obtain
\begin{equation}
\tau_{co} = 4 \tau_V^2 \frac{1}{M}\sum_{q=1}^{M-1} \frac{\delta-\epsilon_q}{(\delta-\epsilon_q)^2+\gamma^2} e^{i\phi^{AC}_q} ,
\label{eqS:long discrete 2}
\end{equation}
for the cotunneling, with the phases $\phi^{AC}_q=d q /M$, and 
\begin{equation}
\hbar \Gamma = 8 \tau_V^2 \frac{1}{M}\sum_{q=1}^{M-1} \frac{\gamma}{(\delta-\epsilon_q)^2+\gamma^2},
\label{eqS:leakage 2}
\end{equation}
for the leakage. It is worth to note the difference of these two formulas. While terms contributing to the leakage (incoherent process) are all positive, there are nontrivial phases appearing in the cotunneling, which is a coherent process. The phases depend on the particle correlations in the band and their destructive interference is very general.\cite{basharov2005:JETP} Devising ways to suppress such interference is probably the most important factor in inducing effective long-range cotunneling.  Ref.~\onlinecite{openov2005:S} shows how these correlations for dots close to the boundary can be changed by changing the array boundary conditions. For us it is critical that the virtual band is not cut at the manipulated dots $A$ and $C$, since that would lead to unfavorable properties of these phases and a very strong destructive interference (see below).

To evaluate Eqs.~\eqref{eqS:long discrete 2} and \eqref{eqS:leakage 2}, we take the continuum limit $M\to \infty$ by introducing a continuum variable $\kappa=q/M$ which covers the band by $\kappa \in (0,1)$. We rewrite the expressions using dimensionless parameters $\Delta=\delta/\gamma$ (the detuning in units of decoherence, of order 1 and larger), $g=\gamma/\tau$ (decoherence in units of the bandwidth, much smaller than 1) and $\xi=2\pi d\surd{\gamma/\tau}$ (a parameter related to the dots distance, which can be both smaller and larger than 1). Assuming the bandwidth is much larger than $\gamma$ and $\delta$, we Taylor expand the virtual states energies in the lowest order in $\kappa$ and get
\begin{equation}
\tau_{co} = \frac{4}{\pi}\frac{\tau_V^2}{\tau} \sqrt{\frac{\tau}{\gamma}} \int_0^{\infty} {\rm d}x\,  e^{i \xi x} \frac{\Delta-x^2}{(\Delta-x^2)^2+1} ,
\label{eqS:long continuous}
\end{equation}
and 
\begin{equation}
\hbar \Gamma  = \frac{8}{\pi} \frac{\tau_V^2}{\tau} \sqrt{\frac{\tau}{\gamma}} \int_0^{\infty} {\rm d}x\,  \frac{1}{(\Delta-x^2)^2+1}.
\label{eqS:leakage continuous}
\end{equation}
The integral in the last equation can be calculated analytically with the result
\begin{equation}
 \frac{\pi}{2} (\Delta^2+1)^{-1/4} \sin\left( \frac{1}{2} \arctan \Big(\frac{1}{\Delta}\Big) \right).
 \label{eqS:auxiliary integral}
\end{equation}
Expanding in the leading order in $1/\Delta$ gives Eq.~\eqref{eq:leakage}. 

We evaluate Eq.~\eqref{eqS:long continuous} in two limits. First, if $1/\xi^2 \gg \Delta$, we replace the phase factor by 1. The integral then equals Eq.~\eqref{eqS:auxiliary integral} multiplied by an additional factor $\Delta+\sqrt{\Delta^2+1}$. In the opposite limit, $1/\xi^2 \ll \Delta$, the fast oscillating phase factor acts effectively as a derivative  $-(2i/\xi) \partial_x$, by which we can replace it. The integral then gives $\sim 2\Delta/\xi(\Delta^2+1)$. Expanding again for large $\Delta$, we obtain the two results of Eq.~\eqref{eq:tac long}.

Let us now come back to the condition on the distance from the boundary. The replacement of functions in Eq.~\eqref{eqS:wave functions} by plane waves [Eq.~\eqref{eqS:plane waves}] will not influence the result for the cotunneling if the upper integration limit in Eq.~\eqref{eqS:long continuous}, being the minimum of $\{1/\xi, \surd{\Delta}\}$, corresponds to a wave vector $q_u$ resulting in a large enough phase in the arguments of functions in Eqs.~\eqref{eqS:wave functions} and \eqref{eqS:plane waves}. The condition therefore reads
\begin{equation}
M \lesssim j_\alpha q_u = j_\alpha M \pi \sqrt{g} x_u =  j_\alpha M \pi \sqrt{g} \, {\rm min} \{1/\xi, \sqrt{\delta/\gamma} \},
\end{equation}
from where we get
\begin{equation}
j_\alpha \gtrsim (2/\pi) {\rm max} \{ d, d_0 \},
\label{eqS:condition on edge distance}
\end{equation}
as the requirement on the distance of the manipulated dots $\alpha=A,C$ from the array edges so that Eq.~\eqref{eqS:long continuous} is valid.

If the condition in Eq.~\eqref{eqS:condition on edge distance} is not fulfilled, the cotunneling, unlike the leakage, will be suppressed by an additional factor. As an illustration of such suppression, we consider the case of the two manipulated dots being on the edge of the array, $j_A=1$, $j_C=M-1$. 
Equation \eqref{eqS:long discrete} applies with only a single term contributing in Eq.~\eqref{eqS:tunneling me aux}, as both $A$ and $C$ have only a single neighbor. Using Eq.~\eqref{eqS:wave functions}, we get
\begin{equation}
\tau_{co} = \tau_V^2 \sum_{q=1}^{M-1} \frac{\delta-\epsilon_q}{(\delta-\epsilon_q)^2+\gamma^2} |\Psi_q(1)|^2 (-1)^q,
\label{eqS:long discrete edge}
\end{equation}
being the worst possible case of destructive interference, corresponding to phases $\phi^{AC}_q=\pi q$. To evaluate the sum, we replace the oscillating sign by a derivative $(-1)^q \to -\partial_q$ and get
\begin{equation}
\tau_{co} = -\tau_V^2  \left[ \frac{\delta-\epsilon_q}{(\delta-\epsilon_q)^2+\gamma^2} |\Psi_q(1)|^2 \right]_{q=1}^{q=M-1}.
\label{eqS:long discrete edge 2}
\end{equation}
Assuming the lower band edge detuning is much smaller than the upper one, and expanding the wavefunction in the leading order in $M$ we get
\begin{equation}
\tau_{co} = \tau_V^2  \frac{2\pi^2}{M^3}  \frac{\delta}{\delta^2+\gamma^2}.
\label{eqS:long discrete edge 3}
\end{equation}
The cotunneling is proportional to the photon assisted tunneling matrix element squared, divided by the detuning, with the latter limited from below by the decoherence. This is the same behavior as that of Eq.~\eqref{eq:tac long} for $d>d_0$. However, here there is an additional quadratic suppression with the distance $d\approx M$ on top the linear fall off of Eq.~\eqref{eq:tac long}. 

The leakage rate can be evaluated for this case too. However, instead of giving explicitly the cumbersome formulas, we only state that, since there is no destructive interference for the leakage, the result is qualitatively the same as the one in Eq.~\eqref{eq:leakage}. Then as long as the decoherence of the band states is larger than their separation $\gamma \gtrsim \tau/M$, which is a natural property of a continuous band, we get that the requirement for the leakage to be smaller than the cotunneling requires very large detunings
\begin{equation}
\delta \gtrsim M^{2/3} \tau.
\end{equation}
This strongly limits possibilities for long distance cotunneling mediated by the dot array itself in this configuration, microwaves assisted or not.

\bibliography{/Users/peter/Xfiles/docs/articles/references/quantum_dot}

\begin{thebibliography}{76}%
\makeatletter
\providecommand \@ifxundefined [1]{%
 \@ifx{#1\undefined}
}%
\providecommand \@ifnum [1]{%
 \ifnum #1\expandafter \@firstoftwo
 \else \expandafter \@secondoftwo
 \fi
}%
\providecommand \@ifx [1]{%
 \ifx #1\expandafter \@firstoftwo
 \else \expandafter \@secondoftwo
 \fi
}%
\providecommand \natexlab [1]{#1}%
\providecommand \enquote  [1]{``#1''}%
\providecommand \bibnamefont  [1]{#1}%
\providecommand \bibfnamefont [1]{#1}%
\providecommand \citenamefont [1]{#1}%
\providecommand \href@noop [0]{\@secondoftwo}%
\providecommand \href [0]{\begingroup \@sanitize@url \@href}%
\providecommand \@href[1]{\@@startlink{#1}\@@href}%
\providecommand \@@href[1]{\endgroup#1\@@endlink}%
\providecommand \@sanitize@url [0]{\catcode `\\12\catcode `\$12\catcode
  `\&12\catcode `\#12\catcode `\^12\catcode `\_12\catcode `\%12\relax}%
\providecommand \@@startlink[1]{}%
\providecommand \@@endlink[0]{}%
\providecommand \url  [0]{\begingroup\@sanitize@url \@url }%
\providecommand \@url [1]{\endgroup\@href {#1}{\urlprefix }}%
\providecommand \urlprefix  [0]{URL }%
\providecommand \Eprint [0]{\href }%
\providecommand \doibase [0]{http://dx.doi.org/}%
\providecommand \selectlanguage [0]{\@gobble}%
\providecommand \bibinfo  [0]{\@secondoftwo}%
\providecommand \bibfield  [0]{\@secondoftwo}%
\providecommand \translation [1]{[#1]}%
\providecommand \BibitemOpen [0]{}%
\providecommand \bibitemStop [0]{}%
\providecommand \bibitemNoStop [0]{.\EOS\space}%
\providecommand \EOS [0]{\spacefactor3000\relax}%
\providecommand \BibitemShut  [1]{\csname bibitem#1\endcsname}%
\let\auto@bib@innerbib\@empty
\bibitem [{\citenamefont {Tien}\ and\ \citenamefont
  {Gordon}(1963)}]{tien1963:PR}%
  \BibitemOpen
  \bibfield  {author} {\bibinfo {author} {\bibfnamefont {P.~K.}\ \bibnamefont
  {Tien}}\ and\ \bibinfo {author} {\bibfnamefont {J.~P.}\ \bibnamefont
  {Gordon}},\ }\href@noop {} {\bibfield  {journal} {\bibinfo  {journal} {Phys.
  Rev.}\ }\textbf {\bibinfo {volume} {129}},\ \bibinfo {pages} {647} (\bibinfo
  {year} {1963})}\BibitemShut {NoStop}%
\bibitem [{\citenamefont {Tucker}\ and\ \citenamefont
  {Feldman}(1985)}]{tucker1985:RMP}%
  \BibitemOpen
  \bibfield  {author} {\bibinfo {author} {\bibfnamefont {J.~R.}\ \bibnamefont
  {Tucker}}\ and\ \bibinfo {author} {\bibfnamefont {M.~J.}\ \bibnamefont
  {Feldman}},\ }\href@noop {} {\bibfield  {journal} {\bibinfo  {journal} {Rev.
  Mod. Phys.}\ }\textbf {\bibinfo {volume} {57}},\ \bibinfo {pages} {1055}
  (\bibinfo {year} {1985})}\BibitemShut {NoStop}%
\bibitem [{\citenamefont {Dayem}\ and\ \citenamefont
  {Martin}(1962)}]{dayem1962:PRL}%
  \BibitemOpen
  \bibfield  {author} {\bibinfo {author} {\bibfnamefont {A.~H.}\ \bibnamefont
  {Dayem}}\ and\ \bibinfo {author} {\bibfnamefont {R.~J.}\ \bibnamefont
  {Martin}},\ }\href@noop {} {\bibfield  {journal} {\bibinfo  {journal} {Phys.
  Rev. Lett.}\ }\textbf {\bibinfo {volume} {8}},\ \bibinfo {pages} {246}
  (\bibinfo {year} {1962})}\BibitemShut {NoStop}%
\bibitem [{\citenamefont {Guimar{\~a}es}\ \emph {et~al.}(1993)\citenamefont
  {Guimar{\~a}es}, \citenamefont {Keay}, \citenamefont {Kaminski},
  \citenamefont {{Allen, Jr.}}, \citenamefont {Hopkins}, \citenamefont
  {Gossard}, \citenamefont {Florez},\ and\ \citenamefont
  {Harbison}}]{guimaraes1993:PRL}%
  \BibitemOpen
  \bibfield  {author} {\bibinfo {author} {\bibfnamefont {P.~S.~S.}\
  \bibnamefont {Guimar{\~a}es}}, \bibinfo {author} {\bibfnamefont {B.~J.}\
  \bibnamefont {Keay}}, \bibinfo {author} {\bibfnamefont {J.~P.}\ \bibnamefont
  {Kaminski}}, \bibinfo {author} {\bibfnamefont {S.~J.}\ \bibnamefont {{Allen,
  Jr.}}}, \bibinfo {author} {\bibfnamefont {P.~F.}\ \bibnamefont {Hopkins}},
  \bibinfo {author} {\bibfnamefont {A.~C.}\ \bibnamefont {Gossard}}, \bibinfo
  {author} {\bibfnamefont {L.~T.}\ \bibnamefont {Florez}}, \ and\ \bibinfo
  {author} {\bibfnamefont {J.~P.}\ \bibnamefont {Harbison}},\ }\href@noop {}
  {\bibfield  {journal} {\bibinfo  {journal} {Phys. Rev. Lett.}\ }\textbf
  {\bibinfo {volume} {70}},\ \bibinfo {pages} {3792} (\bibinfo {year}
  {1993})}\BibitemShut {NoStop}%
\bibitem [{\citenamefont {Kouwenhoven}\ \emph {et~al.}(1994)\citenamefont
  {Kouwenhoven}, \citenamefont {Jauhar}, \citenamefont {Orenstein},
  \citenamefont {McEuen}, \citenamefont {Nagamune}, \citenamefont {Motohisa},\
  and\ \citenamefont {Sakaki}}]{kouwenhoven1994:PRL}%
  \BibitemOpen
  \bibfield  {author} {\bibinfo {author} {\bibfnamefont {L.~P.}\ \bibnamefont
  {Kouwenhoven}}, \bibinfo {author} {\bibfnamefont {S.}~\bibnamefont {Jauhar}},
  \bibinfo {author} {\bibfnamefont {J.}~\bibnamefont {Orenstein}}, \bibinfo
  {author} {\bibfnamefont {P.~L.}\ \bibnamefont {McEuen}}, \bibinfo {author}
  {\bibfnamefont {Y.}~\bibnamefont {Nagamune}}, \bibinfo {author}
  {\bibfnamefont {J.}~\bibnamefont {Motohisa}}, \ and\ \bibinfo {author}
  {\bibfnamefont {H.}~\bibnamefont {Sakaki}},\ }\href@noop {} {\bibfield
  {journal} {\bibinfo  {journal} {Phys. Rev. Lett.}\ }\textbf {\bibinfo
  {volume} {73}},\ \bibinfo {pages} {3443} (\bibinfo {year}
  {1994})}\BibitemShut {NoStop}%
\bibitem [{\citenamefont {Fujisawa}\ and\ \citenamefont
  {Tarucha}(1997{\natexlab{a}})}]{fujisawa1997:SM}%
  \BibitemOpen
  \bibfield  {author} {\bibinfo {author} {\bibfnamefont {T.}~\bibnamefont
  {Fujisawa}}\ and\ \bibinfo {author} {\bibfnamefont {S.}~\bibnamefont
  {Tarucha}},\ }\href@noop {} {\bibfield  {journal} {\bibinfo  {journal}
  {Superlattices and microstructures}\ }\textbf {\bibinfo {volume} {21}},\
  \bibinfo {pages} {247} (\bibinfo {year} {1997}{\natexlab{a}})}\BibitemShut
  {NoStop}%
\bibitem [{\citenamefont {Fujisawa}\ and\ \citenamefont
  {Tarucha}(1997{\natexlab{b}})}]{fujisawa1997:JJAP}%
  \BibitemOpen
  \bibfield  {author} {\bibinfo {author} {\bibfnamefont {T.}~\bibnamefont
  {Fujisawa}}\ and\ \bibinfo {author} {\bibfnamefont {S.}~\bibnamefont
  {Tarucha}},\ }\href@noop {} {\bibfield  {journal} {\bibinfo  {journal} {Jpn.
  J. Appl. Phys.}\ }\textbf {\bibinfo {volume} {36}},\ \bibinfo {pages} {4000}
  (\bibinfo {year} {1997}{\natexlab{b}})}\BibitemShut {NoStop}%
\bibitem [{\citenamefont {Oosterkamp}\ \emph {et~al.}(1997)\citenamefont
  {Oosterkamp}, \citenamefont {Kouwenhoven}, \citenamefont {Koolen},
  \citenamefont {{van der Vaart}},\ and\ \citenamefont
  {Harmans}}]{oosterkamp1997:PRL}%
  \BibitemOpen
  \bibfield  {author} {\bibinfo {author} {\bibfnamefont {T.~H.}\ \bibnamefont
  {Oosterkamp}}, \bibinfo {author} {\bibfnamefont {L.~P.}\ \bibnamefont
  {Kouwenhoven}}, \bibinfo {author} {\bibfnamefont {A.~E.~A.}\ \bibnamefont
  {Koolen}}, \bibinfo {author} {\bibfnamefont {N.~C.}\ \bibnamefont {{van der
  Vaart}}}, \ and\ \bibinfo {author} {\bibfnamefont {C.~J. P.~M.}\ \bibnamefont
  {Harmans}},\ }\href@noop {} {\bibfield  {journal} {\bibinfo  {journal} {Phys.
  Rev. Lett.}\ }\textbf {\bibinfo {volume} {78}},\ \bibinfo {pages} {1536}
  (\bibinfo {year} {1997})}\BibitemShut {NoStop}%
\bibitem [{\citenamefont {Blick}\ \emph {et~al.}(1998)\citenamefont {Blick},
  \citenamefont {{van der Weide}}, \citenamefont {Haug},\ and\ \citenamefont
  {Eberl}}]{blick1998:PRL}%
  \BibitemOpen
  \bibfield  {author} {\bibinfo {author} {\bibfnamefont {R.~H.}\ \bibnamefont
  {Blick}}, \bibinfo {author} {\bibfnamefont {D.~W.}\ \bibnamefont {{van der
  Weide}}}, \bibinfo {author} {\bibfnamefont {R.~J.}\ \bibnamefont {Haug}}, \
  and\ \bibinfo {author} {\bibfnamefont {K.}~\bibnamefont {Eberl}},\
  }\href@noop {} {\bibfield  {journal} {\bibinfo  {journal} {Phys. Rev. Lett.}\
  }\textbf {\bibinfo {volume} {81}},\ \bibinfo {pages} {689} (\bibinfo {year}
  {1998})}\BibitemShut {NoStop}%
\bibitem [{\citenamefont {{van der Wiel}}\ \emph {et~al.}(1999)\citenamefont
  {{van der Wiel}}, \citenamefont {Oosterkamp}, \citenamefont {{De
  Franceschi}}, \citenamefont {Harmans},\ and\ \citenamefont
  {Kouwenhoven}}]{wiel1999:CM}%
  \BibitemOpen
  \bibfield  {author} {\bibinfo {author} {\bibfnamefont {W.}~\bibnamefont {{van
  der Wiel}}}, \bibinfo {author} {\bibfnamefont {T.}~\bibnamefont
  {Oosterkamp}}, \bibinfo {author} {\bibfnamefont {S.}~\bibnamefont {{De
  Franceschi}}}, \bibinfo {author} {\bibfnamefont {C.}~\bibnamefont {Harmans}},
  \ and\ \bibinfo {author} {\bibfnamefont {L.}~\bibnamefont {Kouwenhoven}},\
  }\href@noop {} {\bibfield  {journal} {\bibinfo  {journal} {cond-mat/9904359}\
  } (\bibinfo {year} {1999})}\BibitemShut {NoStop}%
\bibitem [{\citenamefont {Qin}\ \emph {et~al.}(2001)\citenamefont {Qin},
  \citenamefont {Holleitner}, \citenamefont {Eberl},\ and\ \citenamefont
  {Blick}}]{qin2001:PRB}%
  \BibitemOpen
  \bibfield  {author} {\bibinfo {author} {\bibfnamefont {H.}~\bibnamefont
  {Qin}}, \bibinfo {author} {\bibfnamefont {A.~W.}\ \bibnamefont {Holleitner}},
  \bibinfo {author} {\bibfnamefont {K.}~\bibnamefont {Eberl}}, \ and\ \bibinfo
  {author} {\bibfnamefont {R.~H.}\ \bibnamefont {Blick}},\ }\href@noop {}
  {\bibfield  {journal} {\bibinfo  {journal} {Phys. Rev. B}\ }\textbf {\bibinfo
  {volume} {64}},\ \bibinfo {pages} {241302(R)} (\bibinfo {year}
  {2001})}\BibitemShut {NoStop}%
\bibitem [{\citenamefont {Loss}\ and\ \citenamefont
  {DiVincenzo}(1998)}]{loss1998:PRA}%
  \BibitemOpen
  \bibfield  {author} {\bibinfo {author} {\bibfnamefont {D.}~\bibnamefont
  {Loss}}\ and\ \bibinfo {author} {\bibfnamefont {D.~P.}\ \bibnamefont
  {DiVincenzo}},\ }\href@noop {} {\bibfield  {journal} {\bibinfo  {journal}
  {Phys. Rev. A}\ }\textbf {\bibinfo {volume} {57}},\ \bibinfo {pages} {120}
  (\bibinfo {year} {1998})}\BibitemShut {NoStop}%
\bibitem [{\citenamefont {Geerligs}\ \emph {et~al.}(1990)\citenamefont
  {Geerligs}, \citenamefont {Anderegg}, \citenamefont {Holweg}, \citenamefont
  {Mooij}, \citenamefont {Pothier}, \citenamefont {Esteve}, \citenamefont
  {Urbina},\ and\ \citenamefont {Devoret}}]{geerligs1990:PRL}%
  \BibitemOpen
  \bibfield  {author} {\bibinfo {author} {\bibfnamefont {L.~J.}\ \bibnamefont
  {Geerligs}}, \bibinfo {author} {\bibfnamefont {V.~F.}\ \bibnamefont
  {Anderegg}}, \bibinfo {author} {\bibfnamefont {P.~A.~M.}\ \bibnamefont
  {Holweg}}, \bibinfo {author} {\bibfnamefont {J.~E.}\ \bibnamefont {Mooij}},
  \bibinfo {author} {\bibfnamefont {H.}~\bibnamefont {Pothier}}, \bibinfo
  {author} {\bibfnamefont {D.}~\bibnamefont {Esteve}}, \bibinfo {author}
  {\bibfnamefont {C.}~\bibnamefont {Urbina}}, \ and\ \bibinfo {author}
  {\bibfnamefont {M.~H.}\ \bibnamefont {Devoret}},\ }\href@noop {} {\bibfield
  {journal} {\bibinfo  {journal} {Phys. Rev. Lett.}\ }\textbf {\bibinfo
  {volume} {64}},\ \bibinfo {pages} {2691} (\bibinfo {year}
  {1990})}\BibitemShut {NoStop}%
\bibitem [{\citenamefont {Kouwenhoven}\ \emph {et~al.}(1991)\citenamefont
  {Kouwenhoven}, \citenamefont {Johnson}, \citenamefont {{van der Vaart}},
  \citenamefont {Harmans},\ and\ \citenamefont {Foxon}}]{kouwenhoven1991:PRL}%
  \BibitemOpen
  \bibfield  {author} {\bibinfo {author} {\bibfnamefont {L.~P.}\ \bibnamefont
  {Kouwenhoven}}, \bibinfo {author} {\bibfnamefont {A.~T.}\ \bibnamefont
  {Johnson}}, \bibinfo {author} {\bibfnamefont {N.~C.}\ \bibnamefont {{van der
  Vaart}}}, \bibinfo {author} {\bibfnamefont {C.~J. P.~M.}\ \bibnamefont
  {Harmans}}, \ and\ \bibinfo {author} {\bibfnamefont {C.~T.}\ \bibnamefont
  {Foxon}},\ }\href@noop {} {\bibfield  {journal} {\bibinfo  {journal} {Phys.
  Rev. Lett.}\ }\textbf {\bibinfo {volume} {67}},\ \bibinfo {pages} {1626}
  (\bibinfo {year} {1991})}\BibitemShut {NoStop}%
\bibitem [{\citenamefont {Watson}\ \emph {et~al.}(2003)\citenamefont {Watson},
  \citenamefont {Potok}, \citenamefont {Marcus},\ and\ \citenamefont
  {Umansky}}]{watson2003:PRL}%
  \BibitemOpen
  \bibfield  {author} {\bibinfo {author} {\bibfnamefont {S.~K.}\ \bibnamefont
  {Watson}}, \bibinfo {author} {\bibfnamefont {R.~M.}\ \bibnamefont {Potok}},
  \bibinfo {author} {\bibfnamefont {C.~M.}\ \bibnamefont {Marcus}}, \ and\
  \bibinfo {author} {\bibfnamefont {V.}~\bibnamefont {Umansky}},\ }\href@noop
  {} {\bibfield  {journal} {\bibinfo  {journal} {Phys. Rev. Lett.}\ }\textbf
  {\bibinfo {volume} {91}},\ \bibinfo {pages} {258301} (\bibinfo {year}
  {2003})}\BibitemShut {NoStop}%
\bibitem [{\citenamefont {Kim}\ \emph {et~al.}(1999)\citenamefont {Kim},
  \citenamefont {Benson}, \citenamefont {Kan},\ and\ \citenamefont
  {Yamamoto}}]{kim1999:N}%
  \BibitemOpen
  \bibfield  {author} {\bibinfo {author} {\bibfnamefont {J.}~\bibnamefont
  {Kim}}, \bibinfo {author} {\bibfnamefont {O.}~\bibnamefont {Benson}},
  \bibinfo {author} {\bibfnamefont {H.}~\bibnamefont {Kan}}, \ and\ \bibinfo
  {author} {\bibfnamefont {Y.}~\bibnamefont {Yamamoto}},\ }\href@noop {}
  {\bibfield  {journal} {\bibinfo  {journal} {Nature}\ }\textbf {\bibinfo
  {volume} {397}},\ \bibinfo {pages} {500} (\bibinfo {year}
  {1999})}\BibitemShut {NoStop}%
\bibitem [{\citenamefont {Brouwer}(1998)}]{brouwer1998:PRB}%
  \BibitemOpen
  \bibfield  {author} {\bibinfo {author} {\bibfnamefont {P.~W.}\ \bibnamefont
  {Brouwer}},\ }\href@noop {} {\bibfield  {journal} {\bibinfo  {journal} {Phys.
  Rev. B}\ }\textbf {\bibinfo {volume} {58}},\ \bibinfo {pages} {R10135}
  (\bibinfo {year} {1998})}\BibitemShut {NoStop}%
\bibitem [{\citenamefont {Switkes}\ \emph {et~al.}(1999)\citenamefont
  {Switkes}, \citenamefont {Marcus}, \citenamefont {Campman},\ and\
  \citenamefont {Gossard}}]{switkes1999:S}%
  \BibitemOpen
  \bibfield  {author} {\bibinfo {author} {\bibfnamefont {M.}~\bibnamefont
  {Switkes}}, \bibinfo {author} {\bibfnamefont {C.~M.}\ \bibnamefont {Marcus}},
  \bibinfo {author} {\bibfnamefont {K.}~\bibnamefont {Campman}}, \ and\
  \bibinfo {author} {\bibfnamefont {A.~C.}\ \bibnamefont {Gossard}},\
  }\href@noop {} {\bibfield  {journal} {\bibinfo  {journal} {Science}\ }\textbf
  {\bibinfo {volume} {283}},\ \bibinfo {pages} {1905} (\bibinfo {year}
  {1999})}\BibitemShut {NoStop}%
\bibitem [{\citenamefont {Greentree}\ \emph {et~al.}(2004)\citenamefont
  {Greentree}, \citenamefont {Cole}, \citenamefont {Hamilton},\ and\
  \citenamefont {Hollenberg}}]{greentree2004:PRB}%
  \BibitemOpen
  \bibfield  {author} {\bibinfo {author} {\bibfnamefont {A.~D.}\ \bibnamefont
  {Greentree}}, \bibinfo {author} {\bibfnamefont {J.~H.}\ \bibnamefont {Cole}},
  \bibinfo {author} {\bibfnamefont {A.~R.}\ \bibnamefont {Hamilton}}, \ and\
  \bibinfo {author} {\bibfnamefont {L.~C.~L.}\ \bibnamefont {Hollenberg}},\
  }\href@noop {} {\bibfield  {journal} {\bibinfo  {journal} {Phys. Rev. B}\
  }\textbf {\bibinfo {volume} {70}},\ \bibinfo {pages} {235317} (\bibinfo
  {year} {2004})}\BibitemShut {NoStop}%
\bibitem [{\citenamefont {Hohenester}\ \emph {et~al.}(2006)\citenamefont
  {Hohenester}, \citenamefont {Fabian},\ and\ \citenamefont
  {Troiani}}]{hohenester2006:OC}%
  \BibitemOpen
  \bibfield  {author} {\bibinfo {author} {\bibfnamefont {U.}~\bibnamefont
  {Hohenester}}, \bibinfo {author} {\bibfnamefont {J.}~\bibnamefont {Fabian}},
  \ and\ \bibinfo {author} {\bibfnamefont {F.}~\bibnamefont {Troiani}},\
  }\href@noop {} {\bibfield  {journal} {\bibinfo  {journal} {Opt. Comm.}\
  }\textbf {\bibinfo {volume} {264}},\ \bibinfo {pages} {426} (\bibinfo {year}
  {2006})}\BibitemShut {NoStop}%
\bibitem [{\citenamefont {Huneke}\ \emph {et~al.}(2013)\citenamefont {Huneke},
  \citenamefont {Platero},\ and\ \citenamefont {Kohler}}]{huneke2013:PRL}%
  \BibitemOpen
  \bibfield  {author} {\bibinfo {author} {\bibfnamefont {J.}~\bibnamefont
  {Huneke}}, \bibinfo {author} {\bibfnamefont {G.}~\bibnamefont {Platero}}, \
  and\ \bibinfo {author} {\bibfnamefont {S.}~\bibnamefont {Kohler}},\
  }\href@noop {} {\bibfield  {journal} {\bibinfo  {journal} {Phys. Rev. Lett.}\
  }\textbf {\bibinfo {volume} {110}},\ \bibinfo {pages} {036802} (\bibinfo
  {year} {2013})}\BibitemShut {NoStop}%
\bibitem [{\citenamefont {Bruder}\ and\ \citenamefont
  {Schoeller}(1994)}]{bruder1994:PRL}%
  \BibitemOpen
  \bibfield  {author} {\bibinfo {author} {\bibfnamefont {C.}~\bibnamefont
  {Bruder}}\ and\ \bibinfo {author} {\bibfnamefont {H.}~\bibnamefont
  {Schoeller}},\ }\href@noop {} {\bibfield  {journal} {\bibinfo  {journal}
  {Phys. Rev. Lett.}\ }\textbf {\bibinfo {volume} {72}},\ \bibinfo {pages}
  {1076} (\bibinfo {year} {1994})}\BibitemShut {NoStop}%
\bibitem [{\citenamefont {Stafford}\ and\ \citenamefont
  {Wingreen}(1996)}]{stafford1996:PRL}%
  \BibitemOpen
  \bibfield  {author} {\bibinfo {author} {\bibfnamefont {C.~A.}\ \bibnamefont
  {Stafford}}\ and\ \bibinfo {author} {\bibfnamefont {N.~S.}\ \bibnamefont
  {Wingreen}},\ }\href@noop {} {\bibfield  {journal} {\bibinfo  {journal}
  {Phys. Rev. Lett.}\ }\textbf {\bibinfo {volume} {76}},\ \bibinfo {pages}
  {1916} (\bibinfo {year} {1996})}\BibitemShut {NoStop}%
\bibitem [{\citenamefont {Stoof}\ and\ \citenamefont
  {Nazarov}(1996)}]{stoof1996:PRB}%
  \BibitemOpen
  \bibfield  {author} {\bibinfo {author} {\bibfnamefont {T.~H.}\ \bibnamefont
  {Stoof}}\ and\ \bibinfo {author} {\bibfnamefont {Y.~V.}\ \bibnamefont
  {Nazarov}},\ }\href@noop {} {\bibfield  {journal} {\bibinfo  {journal} {Phys.
  Rev. B}\ }\textbf {\bibinfo {volume} {53}},\ \bibinfo {pages} {1050}
  (\bibinfo {year} {1996})}\BibitemShut {NoStop}%
\bibitem [{\citenamefont {Pedersen}\ and\ \citenamefont
  {B{\"u}ttiker}(1998)}]{pedersen1998:PRB}%
  \BibitemOpen
  \bibfield  {author} {\bibinfo {author} {\bibfnamefont {M.~H.}\ \bibnamefont
  {Pedersen}}\ and\ \bibinfo {author} {\bibfnamefont {M.}~\bibnamefont
  {B{\"u}ttiker}},\ }\href@noop {} {\bibfield  {journal} {\bibinfo  {journal}
  {Phys. Rev. B}\ }\textbf {\bibinfo {volume} {58}},\ \bibinfo {pages} {12993}
  (\bibinfo {year} {1998})}\BibitemShut {NoStop}%
\bibitem [{\citenamefont {Hazelzet}\ \emph {et~al.}(2001)\citenamefont
  {Hazelzet}, \citenamefont {Wegewijs}, \citenamefont {Stoof},\ and\
  \citenamefont {Nazarov}}]{hazelzet2001:PRB}%
  \BibitemOpen
  \bibfield  {author} {\bibinfo {author} {\bibfnamefont {B.~L.}\ \bibnamefont
  {Hazelzet}}, \bibinfo {author} {\bibfnamefont {M.~R.}\ \bibnamefont
  {Wegewijs}}, \bibinfo {author} {\bibfnamefont {T.~H.}\ \bibnamefont {Stoof}},
  \ and\ \bibinfo {author} {\bibfnamefont {Y.~V.}\ \bibnamefont {Nazarov}},\
  }\href@noop {} {\bibfield  {journal} {\bibinfo  {journal} {Phys. Rev. B}\
  }\textbf {\bibinfo {volume} {63}},\ \bibinfo {pages} {165313} (\bibinfo
  {year} {2001})}\BibitemShut {NoStop}%
\bibitem [{\citenamefont {Oosterkamp}\ \emph {et~al.}(1998)\citenamefont
  {Oosterkamp}, \citenamefont {Fujisawa}, \citenamefont {van~der Wiel},
  \citenamefont {Ishibashi}, \citenamefont {Hijman}, \citenamefont {Tarucha},\
  and\ \citenamefont {Kouwenhoven}}]{oosterkamp1998:N}%
  \BibitemOpen
  \bibfield  {author} {\bibinfo {author} {\bibfnamefont {T.~H.}\ \bibnamefont
  {Oosterkamp}}, \bibinfo {author} {\bibfnamefont {T.}~\bibnamefont
  {Fujisawa}}, \bibinfo {author} {\bibfnamefont {W.~G.}\ \bibnamefont {van~der
  Wiel}}, \bibinfo {author} {\bibfnamefont {K.}~\bibnamefont {Ishibashi}},
  \bibinfo {author} {\bibfnamefont {R.~V.}\ \bibnamefont {Hijman}}, \bibinfo
  {author} {\bibfnamefont {S.}~\bibnamefont {Tarucha}}, \ and\ \bibinfo
  {author} {\bibfnamefont {L.~P.}\ \bibnamefont {Kouwenhoven}},\ }\href@noop {}
  {\bibfield  {journal} {\bibinfo  {journal} {Nature}\ }\textbf {\bibinfo
  {volume} {395}},\ \bibinfo {pages} {873} (\bibinfo {year}
  {1998})}\BibitemShut {NoStop}%
\bibitem [{\citenamefont {Forster}\ \emph {et~al.}(2014)\citenamefont
  {Forster}, \citenamefont {Petersen}, \citenamefont {Manus}, \citenamefont
  {H{\"a}nggi}, \citenamefont {Schuh}, \citenamefont {Wegscheider},
  \citenamefont {Kohler},\ and\ \citenamefont {Ludwig}}]{forster2014:PRL}%
  \BibitemOpen
  \bibfield  {author} {\bibinfo {author} {\bibfnamefont {F.}~\bibnamefont
  {Forster}}, \bibinfo {author} {\bibfnamefont {G.}~\bibnamefont {Petersen}},
  \bibinfo {author} {\bibfnamefont {S.}~\bibnamefont {Manus}}, \bibinfo
  {author} {\bibfnamefont {P.}~\bibnamefont {H{\"a}nggi}}, \bibinfo {author}
  {\bibfnamefont {D.}~\bibnamefont {Schuh}}, \bibinfo {author} {\bibfnamefont
  {W.}~\bibnamefont {Wegscheider}}, \bibinfo {author} {\bibfnamefont
  {S.}~\bibnamefont {Kohler}}, \ and\ \bibinfo {author} {\bibfnamefont
  {S.}~\bibnamefont {Ludwig}},\ }\href@noop {} {\bibfield  {journal} {\bibinfo
  {journal} {Phys. Rev. Lett.}\ }\textbf {\bibinfo {volume} {112}},\ \bibinfo
  {pages} {116803} (\bibinfo {year} {2014})}\BibitemShut {NoStop}%
\bibitem [{\citenamefont {Grifoni}\ and\ \citenamefont
  {H{\"a}nggi}(1998)}]{grifoni1998:PR}%
  \BibitemOpen
  \bibfield  {author} {\bibinfo {author} {\bibfnamefont {M.}~\bibnamefont
  {Grifoni}}\ and\ \bibinfo {author} {\bibfnamefont {P.}~\bibnamefont
  {H{\"a}nggi}},\ }\href@noop {} {\bibfield  {journal} {\bibinfo  {journal}
  {Phys. Rep.}\ }\textbf {\bibinfo {volume} {304}},\ \bibinfo {pages} {229}
  (\bibinfo {year} {1998})}\BibitemShut {NoStop}%
\bibitem [{\citenamefont {Platero}\ and\ \citenamefont
  {Aguado}(2004)}]{platero2004:PR}%
  \BibitemOpen
  \bibfield  {author} {\bibinfo {author} {\bibfnamefont {G.}~\bibnamefont
  {Platero}}\ and\ \bibinfo {author} {\bibfnamefont {R.}~\bibnamefont
  {Aguado}},\ }\href@noop {} {\bibfield  {journal} {\bibinfo  {journal} {Phys.
  Rep.}\ }\textbf {\bibinfo {volume} {395}},\ \bibinfo {pages} {1} (\bibinfo
  {year} {2004})}\BibitemShut {NoStop}%
\bibitem [{\citenamefont {Hanson}\ \emph {et~al.}(2007)\citenamefont {Hanson},
  \citenamefont {Kouwenhoven}, \citenamefont {Petta}, \citenamefont {Tarucha},\
  and\ \citenamefont {Vandersypen}}]{hanson2007:RMP}%
  \BibitemOpen
  \bibfield  {author} {\bibinfo {author} {\bibfnamefont {R.}~\bibnamefont
  {Hanson}}, \bibinfo {author} {\bibfnamefont {L.~P.}\ \bibnamefont
  {Kouwenhoven}}, \bibinfo {author} {\bibfnamefont {J.~R.}\ \bibnamefont
  {Petta}}, \bibinfo {author} {\bibfnamefont {S.}~\bibnamefont {Tarucha}}, \
  and\ \bibinfo {author} {\bibfnamefont {L.~M.~K.}\ \bibnamefont
  {Vandersypen}},\ }\href@noop {} {\bibfield  {journal} {\bibinfo  {journal}
  {Rev. Mod. Phys.}\ }\textbf {\bibinfo {volume} {79}},\ \bibinfo {pages}
  {1217} (\bibinfo {year} {2007})}\BibitemShut {NoStop}%
\bibitem [{\citenamefont {Kloeffel}\ and\ \citenamefont
  {Loss}(2013)}]{kloeffel2013:ARCMP}%
  \BibitemOpen
  \bibfield  {author} {\bibinfo {author} {\bibfnamefont {C.}~\bibnamefont
  {Kloeffel}}\ and\ \bibinfo {author} {\bibfnamefont {D.}~\bibnamefont
  {Loss}},\ }\href@noop {} {\bibfield  {journal} {\bibinfo  {journal} {Annu.
  Rev. Cond. Mat. Phys.}\ }\textbf {\bibinfo {volume} {4}},\ \bibinfo {pages}
  {51} (\bibinfo {year} {2013})}\BibitemShut {NoStop}%
\bibitem [{\citenamefont {Amaha}\ \emph {et~al.}(2012)\citenamefont {Amaha},
  \citenamefont {Hatano}, \citenamefont {Tamura}, \citenamefont {Teraoka},
  \citenamefont {Kubo}, \citenamefont {Tokura}, \citenamefont {Austing},\ and\
  \citenamefont {Tarucha}}]{amaha2012:PRB}%
  \BibitemOpen
  \bibfield  {author} {\bibinfo {author} {\bibfnamefont {S.}~\bibnamefont
  {Amaha}}, \bibinfo {author} {\bibfnamefont {T.}~\bibnamefont {Hatano}},
  \bibinfo {author} {\bibfnamefont {H.}~\bibnamefont {Tamura}}, \bibinfo
  {author} {\bibfnamefont {S.}~\bibnamefont {Teraoka}}, \bibinfo {author}
  {\bibfnamefont {T.}~\bibnamefont {Kubo}}, \bibinfo {author} {\bibfnamefont
  {Y.}~\bibnamefont {Tokura}}, \bibinfo {author} {\bibfnamefont {D.~G.}\
  \bibnamefont {Austing}}, \ and\ \bibinfo {author} {\bibfnamefont
  {S.}~\bibnamefont {Tarucha}},\ }\href@noop {} {\bibfield  {journal} {\bibinfo
   {journal} {Phys. Rev. B}\ }\textbf {\bibinfo {volume} {85}},\ \bibinfo
  {pages} {081301(R)} (\bibinfo {year} {2012})}\BibitemShut {NoStop}%
\bibitem [{\citenamefont {Delbecq}\ \emph {et~al.}(2014)\citenamefont
  {Delbecq}, \citenamefont {Nakajima}, \citenamefont {Otsuka}, \citenamefont
  {Amaha}, \citenamefont {Watson}, \citenamefont {Manfra},\ and\ \citenamefont
  {Tarucha}}]{delbecq2014:APL}%
  \BibitemOpen
  \bibfield  {author} {\bibinfo {author} {\bibfnamefont {M.~R.}\ \bibnamefont
  {Delbecq}}, \bibinfo {author} {\bibfnamefont {T.}~\bibnamefont {Nakajima}},
  \bibinfo {author} {\bibfnamefont {T.}~\bibnamefont {Otsuka}}, \bibinfo
  {author} {\bibfnamefont {S.}~\bibnamefont {Amaha}}, \bibinfo {author}
  {\bibfnamefont {J.~D.}\ \bibnamefont {Watson}}, \bibinfo {author}
  {\bibfnamefont {M.~J.}\ \bibnamefont {Manfra}}, \ and\ \bibinfo {author}
  {\bibfnamefont {S.}~\bibnamefont {Tarucha}},\ }\href@noop {} {\bibfield
  {journal} {\bibinfo  {journal} {Appl. Phys. Lett.}\ }\textbf {\bibinfo
  {volume} {104}},\ \bibinfo {pages} {183111} (\bibinfo {year}
  {2014})}\BibitemShut {NoStop}%
\bibitem [{\citenamefont {Takakura}\ \emph {et~al.}(2014)\citenamefont
  {Takakura}, \citenamefont {Noiri}, \citenamefont {Obata}, \citenamefont
  {Otsuka}, \citenamefont {Yoneda}, \citenamefont {Yoshida},\ and\
  \citenamefont {Tarucha}}]{takakura2014:APL}%
  \BibitemOpen
  \bibfield  {author} {\bibinfo {author} {\bibfnamefont {T.}~\bibnamefont
  {Takakura}}, \bibinfo {author} {\bibfnamefont {A.}~\bibnamefont {Noiri}},
  \bibinfo {author} {\bibfnamefont {T.}~\bibnamefont {Obata}}, \bibinfo
  {author} {\bibfnamefont {T.}~\bibnamefont {Otsuka}}, \bibinfo {author}
  {\bibfnamefont {J.}~\bibnamefont {Yoneda}}, \bibinfo {author} {\bibfnamefont
  {K.}~\bibnamefont {Yoshida}}, \ and\ \bibinfo {author} {\bibfnamefont
  {S.}~\bibnamefont {Tarucha}},\ }\href@noop {} {\bibfield  {journal} {\bibinfo
   {journal} {Appl. Phys. Lett.}\ }\textbf {\bibinfo {volume} {104}},\ \bibinfo
  {pages} {113109} (\bibinfo {year} {2014})}\BibitemShut {NoStop}%
\bibitem [{\citenamefont {S\'{a}nchez}\ \emph {et~al.}(2014)\citenamefont
  {S\'{a}nchez}, \citenamefont {Granger}, \citenamefont {Gaudreau},
  \citenamefont {Kam}, \citenamefont {{Pioro-Ladri\'{e}re}}, \citenamefont
  {Studenikin}, \citenamefont {Zawadzki}, \citenamefont {Sachrajda},\ and\
  \citenamefont {Platero}}]{sanchez2014:PRL}%
  \BibitemOpen
  \bibfield  {author} {\bibinfo {author} {\bibfnamefont {R.}~\bibnamefont
  {S\'{a}nchez}}, \bibinfo {author} {\bibfnamefont {G.}~\bibnamefont
  {Granger}}, \bibinfo {author} {\bibfnamefont {L.}~\bibnamefont {Gaudreau}},
  \bibinfo {author} {\bibfnamefont {A.}~\bibnamefont {Kam}}, \bibinfo {author}
  {\bibfnamefont {M.}~\bibnamefont {{Pioro-Ladri\'{e}re}}}, \bibinfo {author}
  {\bibfnamefont {S.~A.}\ \bibnamefont {Studenikin}}, \bibinfo {author}
  {\bibfnamefont {P.}~\bibnamefont {Zawadzki}}, \bibinfo {author}
  {\bibfnamefont {A.~S.}\ \bibnamefont {Sachrajda}}, \ and\ \bibinfo {author}
  {\bibfnamefont {G.}~\bibnamefont {Platero}},\ }\href@noop {} {\bibfield
  {journal} {\bibinfo  {journal} {Phys. Rev. Lett.}\ }\textbf {\bibinfo
  {volume} {112}},\ \bibinfo {pages} {176803} (\bibinfo {year}
  {2014})}\BibitemShut {NoStop}%
\bibitem [{\citenamefont {Busl}\ \emph {et~al.}(2013)\citenamefont {Busl},
  \citenamefont {Granger}, \citenamefont {Gaudreau}, \citenamefont
  {{S\'{a}nchez}}, \citenamefont {Kam}, \citenamefont {{Pioro-Ladri\`{e}re}},
  \citenamefont {Studenikin}, \citenamefont {Zawadzki}, \citenamefont
  {Wasilewski}, \citenamefont {Sachrajda},\ and\ \citenamefont
  {Platero}}]{busl2013:NN}%
  \BibitemOpen
  \bibfield  {author} {\bibinfo {author} {\bibfnamefont {M.}~\bibnamefont
  {Busl}}, \bibinfo {author} {\bibfnamefont {G.}~\bibnamefont {Granger}},
  \bibinfo {author} {\bibfnamefont {L.}~\bibnamefont {Gaudreau}}, \bibinfo
  {author} {\bibfnamefont {R.}~\bibnamefont {{S\'{a}nchez}}}, \bibinfo {author}
  {\bibfnamefont {A.}~\bibnamefont {Kam}}, \bibinfo {author} {\bibfnamefont
  {M.}~\bibnamefont {{Pioro-Ladri\`{e}re}}}, \bibinfo {author} {\bibfnamefont
  {S.~A.}\ \bibnamefont {Studenikin}}, \bibinfo {author} {\bibfnamefont
  {P.}~\bibnamefont {Zawadzki}}, \bibinfo {author} {\bibfnamefont {Z.~R.}\
  \bibnamefont {Wasilewski}}, \bibinfo {author} {\bibfnamefont {A.~S.}\
  \bibnamefont {Sachrajda}}, \ and\ \bibinfo {author} {\bibfnamefont
  {G.}~\bibnamefont {Platero}},\ }\href@noop {} {\bibfield  {journal} {\bibinfo
   {journal} {Nat. Nanotechnol.}\ }\textbf {\bibinfo {volume} {8}},\ \bibinfo
  {pages} {261} (\bibinfo {year} {2013})}\BibitemShut {NoStop}%
\bibitem [{\citenamefont {Braakman}\ \emph
  {et~al.}(2013{\natexlab{a}})\citenamefont {Braakman}, \citenamefont
  {Barthelemy}, \citenamefont {Reichl}, \citenamefont {Wegscheider},\ and\
  \citenamefont {Vandersypen}}]{braakman2013:NN}%
  \BibitemOpen
  \bibfield  {author} {\bibinfo {author} {\bibfnamefont {F.~R.}\ \bibnamefont
  {Braakman}}, \bibinfo {author} {\bibfnamefont {P.}~\bibnamefont
  {Barthelemy}}, \bibinfo {author} {\bibfnamefont {C.}~\bibnamefont {Reichl}},
  \bibinfo {author} {\bibfnamefont {W.}~\bibnamefont {Wegscheider}}, \ and\
  \bibinfo {author} {\bibfnamefont {L.~M.~K.}\ \bibnamefont {Vandersypen}},\
  }\href@noop {} {\bibfield  {journal} {\bibinfo  {journal} {Nat. Nanotech.}\
  }\textbf {\bibinfo {volume} {8}},\ \bibinfo {pages} {432} (\bibinfo {year}
  {2013}{\natexlab{a}})}\BibitemShut {NoStop}%
\bibitem [{\citenamefont {Flensberg}(1997)}]{flensberg1997:PRB}%
  \BibitemOpen
  \bibfield  {author} {\bibinfo {author} {\bibfnamefont {K.}~\bibnamefont
  {Flensberg}},\ }\href@noop {} {\bibfield  {journal} {\bibinfo  {journal}
  {Phys. Rev. B}\ }\textbf {\bibinfo {volume} {55}},\ \bibinfo {pages} {13118}
  (\bibinfo {year} {1997})}\BibitemShut {NoStop}%
\bibitem [{\citenamefont {Srinivasa}\ \emph {et~al.}(2015)\citenamefont
  {Srinivasa}, \citenamefont {Xu},\ and\ \citenamefont
  {Taylor}}]{srinivasa2015:PRL}%
  \BibitemOpen
  \bibfield  {author} {\bibinfo {author} {\bibfnamefont {V.}~\bibnamefont
  {Srinivasa}}, \bibinfo {author} {\bibfnamefont {H.}~\bibnamefont {Xu}}, \
  and\ \bibinfo {author} {\bibfnamefont {J.~M.}\ \bibnamefont {Taylor}},\
  }\href@noop {} {\bibfield  {journal} {\bibinfo  {journal} {Phys. Rev. Lett.}\
  }\textbf {\bibinfo {volume} {114}},\ \bibinfo {pages} {226803} (\bibinfo
  {year} {2015})}\BibitemShut {NoStop}%
\bibitem [{\citenamefont {Schreiber}\ \emph {et~al.}(2011)\citenamefont
  {Schreiber}, \citenamefont {Braakman}, \citenamefont {Meunier}, \citenamefont
  {Calado}, \citenamefont {Danon}, \citenamefont {Taylor}, \citenamefont
  {Wegscheider},\ and\ \citenamefont {Vandersypen}}]{schreiber2011:NC}%
  \BibitemOpen
  \bibfield  {author} {\bibinfo {author} {\bibfnamefont {L.~R.}\ \bibnamefont
  {Schreiber}}, \bibinfo {author} {\bibfnamefont {F.~R.}\ \bibnamefont
  {Braakman}}, \bibinfo {author} {\bibfnamefont {T.}~\bibnamefont {Meunier}},
  \bibinfo {author} {\bibfnamefont {V.}~\bibnamefont {Calado}}, \bibinfo
  {author} {\bibfnamefont {J.}~\bibnamefont {Danon}}, \bibinfo {author}
  {\bibfnamefont {J.~M.}\ \bibnamefont {Taylor}}, \bibinfo {author}
  {\bibfnamefont {W.}~\bibnamefont {Wegscheider}}, \ and\ \bibinfo {author}
  {\bibfnamefont {L.~M.~K.}\ \bibnamefont {Vandersypen}},\ }\href@noop {}
  {\bibfield  {journal} {\bibinfo  {journal} {Nat. Comm.}\ }\textbf {\bibinfo
  {volume} {2}},\ \bibinfo {pages} {556} (\bibinfo {year} {2011})}\BibitemShut
  {NoStop}%
\bibitem [{\citenamefont {Braakman}\ \emph
  {et~al.}(2013{\natexlab{b}})\citenamefont {Braakman}, \citenamefont
  {Barthelemy}, \citenamefont {Reichl}, \citenamefont {Wegscheider},\ and\
  \citenamefont {Vandersypen}}]{braakman2013:APL}%
  \BibitemOpen
  \bibfield  {author} {\bibinfo {author} {\bibfnamefont {F.~R.}\ \bibnamefont
  {Braakman}}, \bibinfo {author} {\bibfnamefont {P.}~\bibnamefont
  {Barthelemy}}, \bibinfo {author} {\bibfnamefont {C.}~\bibnamefont {Reichl}},
  \bibinfo {author} {\bibfnamefont {W.}~\bibnamefont {Wegscheider}}, \ and\
  \bibinfo {author} {\bibfnamefont {L.~M.~K.}\ \bibnamefont {Vandersypen}},\
  }\href@noop {} {\bibfield  {journal} {\bibinfo  {journal} {Appl. Phys.
  Lett.}\ }\textbf {\bibinfo {volume} {102}},\ \bibinfo {pages} {112110}
  (\bibinfo {year} {2013}{\natexlab{b}})}\BibitemShut {NoStop}%
\bibitem [{\citenamefont {Braakman}\ \emph {et~al.}(2014)\citenamefont
  {Braakman}, \citenamefont {Danon}, \citenamefont {Schreiber}, \citenamefont
  {Wegscheider},\ and\ \citenamefont {Vandersypen}}]{braakman2014:PRB}%
  \BibitemOpen
  \bibfield  {author} {\bibinfo {author} {\bibfnamefont {F.~R.}\ \bibnamefont
  {Braakman}}, \bibinfo {author} {\bibfnamefont {J.}~\bibnamefont {Danon}},
  \bibinfo {author} {\bibfnamefont {L.~R.}\ \bibnamefont {Schreiber}}, \bibinfo
  {author} {\bibfnamefont {W.}~\bibnamefont {Wegscheider}}, \ and\ \bibinfo
  {author} {\bibfnamefont {L.~M.~K.}\ \bibnamefont {Vandersypen}},\ }\href@noop
  {} {\bibfield  {journal} {\bibinfo  {journal} {Phys. Rev. B}\ }\textbf
  {\bibinfo {volume} {89}},\ \bibinfo {pages} {075417} (\bibinfo {year}
  {2014})}\BibitemShut {NoStop}%
\bibitem [{\citenamefont {Shirley}(1965)}]{shirley1965:PR}%
  \BibitemOpen
  \bibfield  {author} {\bibinfo {author} {\bibfnamefont {J.~H.}\ \bibnamefont
  {Shirley}},\ }\href@noop {} {\bibfield  {journal} {\bibinfo  {journal} {Phys.
  Rev.}\ }\textbf {\bibinfo {volume} {138}},\ \bibinfo {pages} {979} (\bibinfo
  {year} {1965})}\BibitemShut {NoStop}%
\bibitem [{Note1()}]{Note1}%
  \BibitemOpen
  \bibinfo {note} {By changing the driving field amplitudes $V_\alpha
  $.}\BibitemShut {Stop}%
\bibitem [{\citenamefont {Taylor}\ \emph {et~al.}(2007)\citenamefont {Taylor},
  \citenamefont {Petta}, \citenamefont {Johnson}, \citenamefont {Yacoby},
  \citenamefont {Marcus},\ and\ \citenamefont {Lukin}}]{taylor2007:PRB}%
  \BibitemOpen
  \bibfield  {author} {\bibinfo {author} {\bibfnamefont {J.~M.}\ \bibnamefont
  {Taylor}}, \bibinfo {author} {\bibfnamefont {J.~R.}\ \bibnamefont {Petta}},
  \bibinfo {author} {\bibfnamefont {A.~C.}\ \bibnamefont {Johnson}}, \bibinfo
  {author} {\bibfnamefont {A.}~\bibnamefont {Yacoby}}, \bibinfo {author}
  {\bibfnamefont {C.~M.}\ \bibnamefont {Marcus}}, \ and\ \bibinfo {author}
  {\bibfnamefont {M.~D.}\ \bibnamefont {Lukin}},\ }\href@noop {} {\bibfield
  {journal} {\bibinfo  {journal} {Phys. Rev. B}\ }\textbf {\bibinfo {volume}
  {76}},\ \bibinfo {pages} {035315} (\bibinfo {year} {2007})}\BibitemShut
  {NoStop}%
\bibitem [{\citenamefont {{Gallego-Marcos}}\ \emph {et~al.}(2015)\citenamefont
  {{Gallego-Marcos}}, \citenamefont {{S\'{a}nchez}},\ and\ \citenamefont
  {Platero}}]{marcos2015:JAP}%
  \BibitemOpen
  \bibfield  {author} {\bibinfo {author} {\bibfnamefont {F.}~\bibnamefont
  {{Gallego-Marcos}}}, \bibinfo {author} {\bibfnamefont {R.}~\bibnamefont
  {{S\'{a}nchez}}}, \ and\ \bibinfo {author} {\bibfnamefont {G.}~\bibnamefont
  {Platero}},\ }\href@noop {} {\bibfield  {journal} {\bibinfo  {journal} {J.
  Appl. Phys.}\ }\textbf {\bibinfo {volume} {117}},\ \bibinfo {pages} {112808}
  (\bibinfo {year} {2015})}\BibitemShut {NoStop}%
\bibitem [{Note2()}]{Note2}%
  \BibitemOpen
  \bibinfo {note} {Therefore, the approximation we did to arrive at
  Eq.~\protect \textup {\hbox {\mathsurround \z@ \protect \normalfont
  (\ignorespaces \ref {eq:tac initial final simplified}\unskip \@@italiccorr
  )}} can not be made now.}\BibitemShut {Stop}%
\bibitem [{Note3()}]{Note3}%
  \BibitemOpen
  \bibinfo {note} {Higher photon contributions, which might accidentally break
  the far-detuning condition for some $n$, are suppressed exponentially
  $J_n(eV/\hbar \omega )\approx (eV/2\hbar \omega )^n/n!$ at weak driving
  $eV\ll \hbar \omega $.}\BibitemShut {Stop}%
\bibitem [{\citenamefont {Kanai}\ \emph {et~al.}(2011)\citenamefont {Kanai},
  \citenamefont {Deacon}, \citenamefont {Takahashi}, \citenamefont {Oiwa},
  \citenamefont {Yoshida}, \citenamefont {Shibata}, \citenamefont {Hirakawa},
  \citenamefont {Tokura},\ and\ \citenamefont {Tarucha}}]{kanai2011:NN}%
  \BibitemOpen
  \bibfield  {author} {\bibinfo {author} {\bibfnamefont {Y.}~\bibnamefont
  {Kanai}}, \bibinfo {author} {\bibfnamefont {R.~S.}\ \bibnamefont {Deacon}},
  \bibinfo {author} {\bibfnamefont {S.}~\bibnamefont {Takahashi}}, \bibinfo
  {author} {\bibfnamefont {A.}~\bibnamefont {Oiwa}}, \bibinfo {author}
  {\bibfnamefont {K.}~\bibnamefont {Yoshida}}, \bibinfo {author} {\bibfnamefont
  {K.}~\bibnamefont {Shibata}}, \bibinfo {author} {\bibfnamefont
  {K.}~\bibnamefont {Hirakawa}}, \bibinfo {author} {\bibfnamefont
  {Y.}~\bibnamefont {Tokura}}, \ and\ \bibinfo {author} {\bibfnamefont
  {S.}~\bibnamefont {Tarucha}},\ }\href@noop {} {\bibfield  {journal} {\bibinfo
   {journal} {Nat. Nanotech.}\ }\textbf {\bibinfo {volume} {6}},\ \bibinfo
  {pages} {511} (\bibinfo {year} {2011})}\BibitemShut {NoStop}%
\bibitem [{\citenamefont {Obata}\ \emph {et~al.}(2012)\citenamefont {Obata},
  \citenamefont {{Pioro-Ladri\'{e}re}}, \citenamefont {Tokura},\ and\
  \citenamefont {Tarucha}}]{obata2012:NJP}%
  \BibitemOpen
  \bibfield  {author} {\bibinfo {author} {\bibfnamefont {T.}~\bibnamefont
  {Obata}}, \bibinfo {author} {\bibfnamefont {M.}~\bibnamefont
  {{Pioro-Ladri\'{e}re}}}, \bibinfo {author} {\bibfnamefont {Y.}~\bibnamefont
  {Tokura}}, \ and\ \bibinfo {author} {\bibfnamefont {S.}~\bibnamefont
  {Tarucha}},\ }\href@noop {} {\bibfield  {journal} {\bibinfo  {journal} {New
  J. of Phys.}\ }\textbf {\bibinfo {volume} {14}},\ \bibinfo {pages} {123013}
  (\bibinfo {year} {2012})}\BibitemShut {NoStop}%
\bibitem [{\citenamefont {Kyriakidis}\ \emph {et~al.}(2002)\citenamefont
  {Kyriakidis}, \citenamefont {{Pioro-Ladri\'{e}re}}, \citenamefont {Ciorga},
  \citenamefont {Sachrajda},\ and\ \citenamefont
  {Hawrylak}}]{kyriakidis2002:PRB}%
  \BibitemOpen
  \bibfield  {author} {\bibinfo {author} {\bibfnamefont {J.}~\bibnamefont
  {Kyriakidis}}, \bibinfo {author} {\bibfnamefont {M.}~\bibnamefont
  {{Pioro-Ladri\'{e}re}}}, \bibinfo {author} {\bibfnamefont {M.}~\bibnamefont
  {Ciorga}}, \bibinfo {author} {\bibfnamefont {A.~S.}\ \bibnamefont
  {Sachrajda}}, \ and\ \bibinfo {author} {\bibfnamefont {P.}~\bibnamefont
  {Hawrylak}},\ }\href@noop {} {\bibfield  {journal} {\bibinfo  {journal}
  {Phys. Rev. B}\ }\textbf {\bibinfo {volume} {66}},\ \bibinfo {pages} {035320}
  (\bibinfo {year} {2002})}\BibitemShut {NoStop}%
\bibitem [{\citenamefont {Petta}\ \emph {et~al.}(2005)\citenamefont {Petta},
  \citenamefont {Johnson}, \citenamefont {Taylor}, \citenamefont {Laird},
  \citenamefont {Yacoby}, \citenamefont {Lukin}, \citenamefont {Marcus},
  \citenamefont {Hanson},\ and\ \citenamefont {Gossard}}]{petta2005:S}%
  \BibitemOpen
  \bibfield  {author} {\bibinfo {author} {\bibfnamefont {J.~R.}\ \bibnamefont
  {Petta}}, \bibinfo {author} {\bibfnamefont {A.~C.}\ \bibnamefont {Johnson}},
  \bibinfo {author} {\bibfnamefont {J.~M.}\ \bibnamefont {Taylor}}, \bibinfo
  {author} {\bibfnamefont {E.~A.}\ \bibnamefont {Laird}}, \bibinfo {author}
  {\bibfnamefont {A.}~\bibnamefont {Yacoby}}, \bibinfo {author} {\bibfnamefont
  {M.~D.}\ \bibnamefont {Lukin}}, \bibinfo {author} {\bibfnamefont {C.~M.}\
  \bibnamefont {Marcus}}, \bibinfo {author} {\bibfnamefont {M.~P.}\
  \bibnamefont {Hanson}}, \ and\ \bibinfo {author} {\bibfnamefont {A.~C.}\
  \bibnamefont {Gossard}},\ }\href@noop {} {\bibfield  {journal} {\bibinfo
  {journal} {Science}\ }\textbf {\bibinfo {volume} {309}},\ \bibinfo {pages}
  {2180} (\bibinfo {year} {2005})}\BibitemShut {NoStop}%
\bibitem [{\citenamefont {Koppens}\ \emph {et~al.}(2008)\citenamefont
  {Koppens}, \citenamefont {Nowack},\ and\ \citenamefont
  {Vandersypen}}]{koppens2008:PRL}%
  \BibitemOpen
  \bibfield  {author} {\bibinfo {author} {\bibfnamefont {F.~H.~L.}\
  \bibnamefont {Koppens}}, \bibinfo {author} {\bibfnamefont {K.~C.}\
  \bibnamefont {Nowack}}, \ and\ \bibinfo {author} {\bibfnamefont {L.~M.~K.}\
  \bibnamefont {Vandersypen}},\ }\href@noop {} {\bibfield  {journal} {\bibinfo
  {journal} {Phys. Rev. Lett.}\ }\textbf {\bibinfo {volume} {100}},\ \bibinfo
  {pages} {236802} (\bibinfo {year} {2008})}\BibitemShut {NoStop}%
\bibitem [{\citenamefont {Bluhm}\ \emph {et~al.}(2010)\citenamefont {Bluhm},
  \citenamefont {Foletti}, \citenamefont {Neder}, \citenamefont {Rudner},
  \citenamefont {Mahalu}, \citenamefont {Umansky},\ and\ \citenamefont
  {Yacoby}}]{bluhm2010:N}%
  \BibitemOpen
  \bibfield  {author} {\bibinfo {author} {\bibfnamefont {H.}~\bibnamefont
  {Bluhm}}, \bibinfo {author} {\bibfnamefont {S.}~\bibnamefont {Foletti}},
  \bibinfo {author} {\bibfnamefont {I.}~\bibnamefont {Neder}}, \bibinfo
  {author} {\bibfnamefont {M.}~\bibnamefont {Rudner}}, \bibinfo {author}
  {\bibfnamefont {D.}~\bibnamefont {Mahalu}}, \bibinfo {author} {\bibfnamefont
  {V.}~\bibnamefont {Umansky}}, \ and\ \bibinfo {author} {\bibfnamefont
  {A.}~\bibnamefont {Yacoby}},\ }\href@noop {} {\bibfield  {journal} {\bibinfo
  {journal} {Nat. Phys.}\ }\textbf {\bibinfo {volume} {7}},\ \bibinfo {pages}
  {109} (\bibinfo {year} {2010})}\BibitemShut {NoStop}%
\bibitem [{\citenamefont {Openov}(1999)}]{openov1999:PRB}%
  \BibitemOpen
  \bibfield  {author} {\bibinfo {author} {\bibfnamefont {L.~A.}\ \bibnamefont
  {Openov}},\ }\href@noop {} {\bibfield  {journal} {\bibinfo  {journal} {Phys.
  Rev. B}\ }\textbf {\bibinfo {volume} {60}},\ \bibinfo {pages} {8798}
  (\bibinfo {year} {1999})}\BibitemShut {NoStop}%
\bibitem [{\citenamefont {Sanchez}\ \emph {et~al.}(2014)\citenamefont
  {Sanchez}, \citenamefont {Gallego-Marcos},\ and\ \citenamefont
  {Platero}}]{sanchez2014:PRB}%
  \BibitemOpen
  \bibfield  {author} {\bibinfo {author} {\bibfnamefont {R.}~\bibnamefont
  {Sanchez}}, \bibinfo {author} {\bibfnamefont {F.}~\bibnamefont
  {Gallego-Marcos}}, \ and\ \bibinfo {author} {\bibfnamefont {G.}~\bibnamefont
  {Platero}},\ }\href@noop {} {\bibfield  {journal} {\bibinfo  {journal} {Phys.
  Rev. B}\ }\textbf {\bibinfo {volume} {89}},\ \bibinfo {pages} {161402(R)}
  (\bibinfo {year} {2014})}\BibitemShut {NoStop}%
\bibitem [{\citenamefont {Lehmann}\ \emph {et~al.}(2007)\citenamefont
  {Lehmann}, \citenamefont {{Gaita-Arin}}, \citenamefont {Coronado},\ and\
  \citenamefont {Loss}}]{lehmann2007:NN}%
  \BibitemOpen
  \bibfield  {author} {\bibinfo {author} {\bibfnamefont {J.}~\bibnamefont
  {Lehmann}}, \bibinfo {author} {\bibfnamefont {A.}~\bibnamefont
  {{Gaita-Arin}}}, \bibinfo {author} {\bibfnamefont {E.}~\bibnamefont
  {Coronado}}, \ and\ \bibinfo {author} {\bibfnamefont {D.}~\bibnamefont
  {Loss}},\ }\href@noop {} {\bibfield  {journal} {\bibinfo  {journal} {Nat.
  Nanotech.}\ }\textbf {\bibinfo {volume} {2}},\ \bibinfo {pages} {312}
  (\bibinfo {year} {2007})}\BibitemShut {NoStop}%
\bibitem [{\citenamefont {Stano}\ and\ \citenamefont
  {Fabian}(2005)}]{stano2005:PRB}%
  \BibitemOpen
  \bibfield  {author} {\bibinfo {author} {\bibfnamefont {P.}~\bibnamefont
  {Stano}}\ and\ \bibinfo {author} {\bibfnamefont {J.}~\bibnamefont {Fabian}},\
  }\href@noop {} {\bibfield  {journal} {\bibinfo  {journal} {Phys. Rev. B}\
  }\textbf {\bibinfo {volume} {72}},\ \bibinfo {pages} {155410} (\bibinfo
  {year} {2005})}\BibitemShut {NoStop}%
\bibitem [{\citenamefont {Raith}\ \emph {et~al.}(2012)\citenamefont {Raith},
  \citenamefont {Stano}, \citenamefont {Baruffa},\ and\ \citenamefont
  {Fabian}}]{raith2012:PRL}%
  \BibitemOpen
  \bibfield  {author} {\bibinfo {author} {\bibfnamefont {M.}~\bibnamefont
  {Raith}}, \bibinfo {author} {\bibfnamefont {P.}~\bibnamefont {Stano}},
  \bibinfo {author} {\bibfnamefont {F.}~\bibnamefont {Baruffa}}, \ and\
  \bibinfo {author} {\bibfnamefont {J.}~\bibnamefont {Fabian}},\ }\href@noop {}
  {\bibfield  {journal} {\bibinfo  {journal} {Phys. Rev. Lett.}\ }\textbf
  {\bibinfo {volume} {108}},\ \bibinfo {pages} {246602} (\bibinfo {year}
  {2012})}\BibitemShut {NoStop}%
\bibitem [{\citenamefont {Nichol}\ \emph {et~al.}(2015)\citenamefont {Nichol},
  \citenamefont {Harvey}, \citenamefont {Shulman}, \citenamefont {Pal},
  \citenamefont {Umansky}, \citenamefont {Rashba}, \citenamefont {Halperin},\
  and\ \citenamefont {Yacoby}}]{nichol2015:CM}%
  \BibitemOpen
  \bibfield  {author} {\bibinfo {author} {\bibfnamefont {J.~M.}\ \bibnamefont
  {Nichol}}, \bibinfo {author} {\bibfnamefont {S.~P.}\ \bibnamefont {Harvey}},
  \bibinfo {author} {\bibfnamefont {M.~D.}\ \bibnamefont {Shulman}}, \bibinfo
  {author} {\bibfnamefont {A.}~\bibnamefont {Pal}}, \bibinfo {author}
  {\bibfnamefont {V.}~\bibnamefont {Umansky}}, \bibinfo {author} {\bibfnamefont
  {E.~I.}\ \bibnamefont {Rashba}}, \bibinfo {author} {\bibfnamefont {B.~I.}\
  \bibnamefont {Halperin}}, \ and\ \bibinfo {author} {\bibfnamefont
  {A.}~\bibnamefont {Yacoby}},\ }\href@noop {} {\bibfield  {journal} {\bibinfo
  {journal} {arxiv:1502:05400}\ } (\bibinfo {year} {2015})}\BibitemShut
  {NoStop}%
\bibitem [{Note4()}]{Note4}%
  \BibitemOpen
  \bibinfo {note} {We mean in the continuum limit: adding a finite imaginary
  part into the denominator of Eq.~\protect \textup {\hbox {\mathsurround \z@
  \protect \normalfont (\ignorespaces \ref {eq:tac long 1}\unskip \@@italiccorr
  )}} the condition on the numerator being much smaller than the denominator is
  for $M\to \infty $ trivially fulfilled for each term of the summation, as the
  numerator is proportional to $1/M$.}\BibitemShut {Stop}%
\bibitem [{Note5()}]{Note5}%
  \BibitemOpen
  \bibinfo {note} {The condition is $j_A,j_C\gtrsim (2/\pi ) {\protect \rm max}
  \protect \{d_0,d\protect \}$, with $d_0$ given below Eq.~\protect \textup
  {\hbox {\mathsurround \z@ \protect \normalfont (\ignorespaces \ref {eq:tac
  long}\unskip \@@italiccorr )}}. For dots closer to the array edges, the
  cotunneling is suppressed, unlike the leakage. For $j_A=1$ and $j_C=M-1$,
  representing the worst case, the cotunneling falls-off as $1/d^3$; see
  Eq.~\protect \textup {\hbox {\mathsurround \z@ \protect \normalfont
  (\ignorespaces \ref {eqS:long discrete edge 3}\unskip \@@italiccorr
  )}}.}\BibitemShut {Stop}%
\bibitem [{\citenamefont {Averin}\ and\ \citenamefont
  {Nazarov}(1990)}]{averin1990:PRL}%
  \BibitemOpen
  \bibfield  {author} {\bibinfo {author} {\bibfnamefont {D.~V.}\ \bibnamefont
  {Averin}}\ and\ \bibinfo {author} {\bibfnamefont {Y.~V.}\ \bibnamefont
  {Nazarov}},\ }\href@noop {} {\bibfield  {journal} {\bibinfo  {journal} {Phys.
  Rev. Lett.}\ }\textbf {\bibinfo {volume} {65}},\ \bibinfo {pages} {2446}
  (\bibinfo {year} {1990})}\BibitemShut {NoStop}%
\bibitem [{\citenamefont {Basharov}\ and\ \citenamefont
  {Dubovis}(2005)}]{basharov2005:JETP}%
  \BibitemOpen
  \bibfield  {author} {\bibinfo {author} {\bibfnamefont {A.~M.}\ \bibnamefont
  {Basharov}}\ and\ \bibinfo {author} {\bibfnamefont {S.~A.}\ \bibnamefont
  {Dubovis}},\ }\href@noop {} {\bibfield  {journal} {\bibinfo  {journal}
  {Journal of Exp. and Theor. Phys.}\ }\textbf {\bibinfo {volume} {101}},\
  \bibinfo {pages} {410} (\bibinfo {year} {2005})}\BibitemShut {NoStop}%
\bibitem [{\citenamefont {Openov}\ and\ \citenamefont
  {Tsukanov}(2005)}]{openov2005:S}%
  \BibitemOpen
  \bibfield  {author} {\bibinfo {author} {\bibfnamefont {L.}~\bibnamefont
  {Openov}}\ and\ \bibinfo {author} {\bibfnamefont {A.~V.}\ \bibnamefont
  {Tsukanov}},\ }\href@noop {} {\bibfield  {journal} {\bibinfo  {journal}
  {Semiconductors}\ }\textbf {\bibinfo {volume} {39}},\ \bibinfo {pages} {235}
  (\bibinfo {year} {2005})}\BibitemShut {NoStop}%
\bibitem [{Note6()}]{Note6}%
  \BibitemOpen
  \bibinfo {note} {An alternative simple derivation comes from defining the
  switching as energy conserving, rather than instantaneous. Then, assuming
  $\tau _q \ll \Delta _q$, the state energies after the switch-on are $-\tau
  _q^2/\Delta _q$, and $\Delta _q+\tau _q^2/\Delta _q$. With the system being
  initially in the ground state and with zero couplings, the energy is
  preserved if at a finite coupling the occupation of the excited state is
  $P_2^\prime =\tau _q^2/(\Delta _q^2+2\tau _q^2)$. Considering that both,
  switching-on and switching-off $\tau _q$, generates the same amount of
  charging probability, we arrive at Eq.~\protect \textup {\hbox {\mathsurround
  \z@ \protect \normalfont (\ignorespaces \ref {eqS:P2}\unskip \@@italiccorr
  )}} within the perturbative order adopted in the derivations.}\BibitemShut
  {Stop}%
\bibitem [{Note7()}]{Note7}%
  \BibitemOpen
  \bibinfo {note} {Focus on an excited state here means to nearly align it in
  energy with the ground state by using a gate potential.}\BibitemShut {Stop}%
\bibitem [{\citenamefont {Ratner}(1990)}]{ratner1990:JPC}%
  \BibitemOpen
  \bibfield  {author} {\bibinfo {author} {\bibfnamefont {M.~A.}\ \bibnamefont
  {Ratner}},\ }\href@noop {} {\bibfield  {journal} {\bibinfo  {journal} {J.
  Phys. Chem.}\ }\textbf {\bibinfo {volume} {94}},\ \bibinfo {pages} {4877}
  (\bibinfo {year} {1990})}\BibitemShut {NoStop}%
\bibitem [{\citenamefont {Hanson}\ \emph {et~al.}(2005)\citenamefont {Hanson},
  \citenamefont {{van Beveren}}, \citenamefont {Vink}, \citenamefont
  {Elzerman}, \citenamefont {Naber}, \citenamefont {Koppens}, \citenamefont
  {Kouwenhoven},\ and\ \citenamefont {Vandersypen}}]{hanson2005:PRL}%
  \BibitemOpen
  \bibfield  {author} {\bibinfo {author} {\bibfnamefont {R.}~\bibnamefont
  {Hanson}}, \bibinfo {author} {\bibfnamefont {L.~H.~W.}\ \bibnamefont {{van
  Beveren}}}, \bibinfo {author} {\bibfnamefont {I.~T.}\ \bibnamefont {Vink}},
  \bibinfo {author} {\bibfnamefont {J.~M.}\ \bibnamefont {Elzerman}}, \bibinfo
  {author} {\bibfnamefont {W.~J.~M.}\ \bibnamefont {Naber}}, \bibinfo {author}
  {\bibfnamefont {F.~H.~L.}\ \bibnamefont {Koppens}}, \bibinfo {author}
  {\bibfnamefont {L.~P.}\ \bibnamefont {Kouwenhoven}}, \ and\ \bibinfo {author}
  {\bibfnamefont {L.~M.~K.}\ \bibnamefont {Vandersypen}},\ }\href@noop {}
  {\bibfield  {journal} {\bibinfo  {journal} {Phys. Rev. Lett.}\ }\textbf
  {\bibinfo {volume} {94}},\ \bibinfo {pages} {196802} (\bibinfo {year}
  {2005})}\BibitemShut {NoStop}%
\bibitem [{\citenamefont {Brunner}\ \emph {et~al.}(2011)\citenamefont
  {Brunner}, \citenamefont {Shin}, \citenamefont {Obata}, \citenamefont
  {{Pioro-Ladri\'{e}re}}, \citenamefont {Kubo}, \citenamefont {Yoshida},
  \citenamefont {Taniyama}, \citenamefont {Tokura},\ and\ \citenamefont
  {Tarucha}}]{brunner2011:PRL}%
  \BibitemOpen
  \bibfield  {author} {\bibinfo {author} {\bibfnamefont {R.}~\bibnamefont
  {Brunner}}, \bibinfo {author} {\bibfnamefont {Y.-S.}\ \bibnamefont {Shin}},
  \bibinfo {author} {\bibfnamefont {T.}~\bibnamefont {Obata}}, \bibinfo
  {author} {\bibfnamefont {M.}~\bibnamefont {{Pioro-Ladri\'{e}re}}}, \bibinfo
  {author} {\bibfnamefont {T.}~\bibnamefont {Kubo}}, \bibinfo {author}
  {\bibfnamefont {K.}~\bibnamefont {Yoshida}}, \bibinfo {author} {\bibfnamefont
  {T.}~\bibnamefont {Taniyama}}, \bibinfo {author} {\bibfnamefont
  {Y.}~\bibnamefont {Tokura}}, \ and\ \bibinfo {author} {\bibfnamefont
  {S.}~\bibnamefont {Tarucha}},\ }\href@noop {} {\bibfield  {journal} {\bibinfo
   {journal} {Phys. Rev. Lett.}\ }\textbf {\bibinfo {volume} {107}},\ \bibinfo
  {pages} {146801} (\bibinfo {year} {2011})}\BibitemShut {NoStop}%
\bibitem [{\citenamefont {Stepanenko}\ \emph {et~al.}(2012)\citenamefont
  {Stepanenko}, \citenamefont {Rudner}, \citenamefont {Halperin},\ and\
  \citenamefont {Loss}}]{stepanenko2012:PRB}%
  \BibitemOpen
  \bibfield  {author} {\bibinfo {author} {\bibfnamefont {D.}~\bibnamefont
  {Stepanenko}}, \bibinfo {author} {\bibfnamefont {M.}~\bibnamefont {Rudner}},
  \bibinfo {author} {\bibfnamefont {B.~I.}\ \bibnamefont {Halperin}}, \ and\
  \bibinfo {author} {\bibfnamefont {D.}~\bibnamefont {Loss}},\ }\href@noop {}
  {\bibfield  {journal} {\bibinfo  {journal} {Phys. Rev. B}\ }\textbf {\bibinfo
  {volume} {85}},\ \bibinfo {pages} {075416} (\bibinfo {year}
  {2012})}\BibitemShut {NoStop}%
\bibitem [{\citenamefont {Erlingsson}\ \emph {et~al.}(2001)\citenamefont
  {Erlingsson}, \citenamefont {Nazarov},\ and\ \citenamefont
  {{Fa\v{l}ko}}}]{erlingsson2001:PRB}%
  \BibitemOpen
  \bibfield  {author} {\bibinfo {author} {\bibfnamefont {S.~I.}\ \bibnamefont
  {Erlingsson}}, \bibinfo {author} {\bibfnamefont {Y.~V.}\ \bibnamefont
  {Nazarov}}, \ and\ \bibinfo {author} {\bibfnamefont {V.~I.}\ \bibnamefont
  {{Fa\v{l}ko}}},\ }\href@noop {} {\bibfield  {journal} {\bibinfo  {journal}
  {Phys. Rev. B}\ }\textbf {\bibinfo {volume} {64}},\ \bibinfo {pages} {195306}
  (\bibinfo {year} {2001})}\BibitemShut {NoStop}%
\bibitem [{\citenamefont {Khaetskii}\ \emph {et~al.}(2002)\citenamefont
  {Khaetskii}, \citenamefont {Loss},\ and\ \citenamefont
  {Glazman}}]{khaetskii2002:PRL}%
  \BibitemOpen
  \bibfield  {author} {\bibinfo {author} {\bibfnamefont {A.~V.}\ \bibnamefont
  {Khaetskii}}, \bibinfo {author} {\bibfnamefont {D.}~\bibnamefont {Loss}}, \
  and\ \bibinfo {author} {\bibfnamefont {L.}~\bibnamefont {Glazman}},\
  }\href@noop {} {\bibfield  {journal} {\bibinfo  {journal} {Phys. Rev. Lett.}\
  }\textbf {\bibinfo {volume} {88}},\ \bibinfo {pages} {186802} (\bibinfo
  {year} {2002})}\BibitemShut {NoStop}%
\bibitem [{\citenamefont {Khaetskii}\ and\ \citenamefont
  {Nazarov}(2000)}]{khaetskii2000:PRB}%
  \BibitemOpen
  \bibfield  {author} {\bibinfo {author} {\bibfnamefont {A.~V.}\ \bibnamefont
  {Khaetskii}}\ and\ \bibinfo {author} {\bibfnamefont {Y.~V.}\ \bibnamefont
  {Nazarov}},\ }\href@noop {} {\bibfield  {journal} {\bibinfo  {journal} {Phys.
  Rev. B}\ }\textbf {\bibinfo {volume} {61}},\ \bibinfo {pages} {12639}
  (\bibinfo {year} {2000})}\BibitemShut {NoStop}%
\bibitem [{\citenamefont {Stano}\ and\ \citenamefont
  {Fabian}(2006)}]{stano2006:PRB}%
  \BibitemOpen
  \bibfield  {author} {\bibinfo {author} {\bibfnamefont {P.}~\bibnamefont
  {Stano}}\ and\ \bibinfo {author} {\bibfnamefont {J.}~\bibnamefont {Fabian}},\
  }\href@noop {} {\bibfield  {journal} {\bibinfo  {journal} {Phys. Rev. B}\
  }\textbf {\bibinfo {volume} {74}},\ \bibinfo {pages} {045320} (\bibinfo
  {year} {2006})}\BibitemShut {NoStop}%
\end{thebibliography}%

\end{document}